\documentclass[preliminary,copyright,creativecommons]{eptcs}
\providecommand{\event}{QPL 2026} 

\usepackage{iftex}
\usepackage{todonotes}
\usepackage{amsmath}
\usepackage{algpseudocode}
\usepackage{algorithm}
\usepackage{caption}
\usepackage{subcaption}
\usepackage{amsthm}
\usepackage{multirow}
\usepackage{comment}
\usepackage{xcolor,colortbl}
\definecolor{Gray}{gray}{0.85}
\newcolumntype{g}{>{\columncolor{Gray}}c}

\ifpdf
  \usepackage{underscore}         
  \usepackage[T1]{fontenc}        
\else
  \usepackage{breakurl}           
\fi

\title{Pauli Gadget Synthesis for Gatesets with Arbitrary Even Arity Clifford Gates}
\author{
    Leo Becker
    \institute{University Of Helsinki\\ Helsinki, Finland}
    \email{leo.becker@helsinki.fi}
    \and
    Arianne Meijer - van de Griend
    \institute{University Of Helsinki\\ Helsinki, Finland}
    \email{ariannemeijer@gmail.com}
}
\def\titlerunning{Pauli Gadget Synthesis for Gatesets with Arbitrary Even Arity Clifford Gates}
\def\authorrunning{Leo Becker \& Arianne Meijer - van de Griend}
\begin{document}
\maketitle

\begin{abstract}

We propose an algorithm that can synthesize circuits to gate sets consisting of a $2n$-arity Clifford gate and arbitrary single qubit rotations. To do this, we represent the circuit as a sequence of Pauli gadgets and jointly synthesize them by placing the multi-qubit Clifford in the synthesized circuit and pushing its adjoint through the remaining gadgets such that some qubits are disconnected from some gadgets. The pushed gadgets are collected in a Clifford tableau, which is decomposed afterwards.

With the help of our algorithm we run experiments where we synthesize circuits to varying sizes of target multi-qubit gates. These experiments indicate that increases in gate size can lead to smaller output circuits, and that different connectivity constraints have meaningful differences in how higher arity gates effect the circuit sized.



\end{abstract}

\section{Introduction}
In the current age of quantum computing, it has not yet been determined what will be the main technology driving future quantum computers. In order to successfully compare different computing technologies, we need the ability to determine how hard it is to run programs on those technologies. For this, we need efficient compilers that lead to optimal use of these computers. However, compilers and quantum computers are implicitly co-designed; quantum computers expose operations for which good compilers exist, and compilers are made to work with those exposed operations. This feedback loop can be detrimental for new technologies that do not directly fit the assumptions made by existing compiler software, as compiling can become very inefficient or outright fail, leading to an unfair representation of the underlying hardware.

For example, current compilers assume that the gate set supported by the quantum computer contains gates acting on at most two qubits. These are the gates that are generally made available on existing devices, but depending on the quantum computer, it might be possible to expose gates with higher arity, for example in superconductors or ion-traps~\cite{roy2020programmable, schindler2013quantum}. In order to facilitate the usage of higher arity gates, we introduce a synthesizing approach similar to PauliSteinerGraySynth~\cite{huang2024redefininglexicographicalorderingoptimizing}, which assumes that the single multi-qubit operator of choice is an even arity Clifford Pauli operator, rather than a CNOT. Note that this is different from works that assume the existence of a multi-qubit operator between arbitrary sets of qubits (e.g.~\cite{villoria2026optimisation}).

Using our synthesizing approach, we run experiments on even multi-qubit gate sizes from two to twenty qubits with varying connectivity constraints. The results from these experiments indicate that increases in gate size can lead to large benefits, and that the benefits or disadvantages, vary between different connectivity constraints.

The benefits from our research are twofold; first, we demonstrate that the synthesis approach can be adapted for different gate sets, and second, our experimental results give an indication of how the size of the program changes with the size of the hardware supported entangling gate.

The structure of this work is as follows. Section~\ref{background} explains background to the latter sections. Then Section~\ref{methods} introduces our algorithm, which is then followed by section~\ref{experiments} that contains some experiments with said algorithm. Finally section section \ref{discussion} contains some discussion and concludes the work.

\section{Background}\label{background}
Our approach for circuit synthesis is based on representing circuits as sequences of Pauli gadgets $e^{i\theta P}$, where $\theta \in \mathbf{R}$ is an angle and $P$ a tensor product of the Pauli matrices $X$, $Y$, $Z$, and the identity matrix $I$. For convenience, the matrices $X$, $Y$, $Z$, and $I$ are called \textit{Pauli letters}, and their tensor product $P$ is called a \textit{Pauli string}. Additionally, we will omit the tensor product symbols from Pauli strings, meaning that for example the Pauli string $X\otimes I\otimes Z\otimes Y\otimes Y$ can be written more concisely as $XIZYY$.

In some cases, we have sequences of Pauli gadgets that are unordered, which allows additional optimization during synthesis, e.g. using phase folding~\cite{matroidpartitioning}, or jointly synthesizing Pauli gadgets that share qubits. These unordered sequences can originate from parts of circuits where every operation commutes, i.e. $e^{i\theta_i P_i}e^{i\theta_j P_j}=e^{i\theta_j P_j}e^{i\theta_i P_i}$ holds for every operation pair in the sequence. Alternatively, if the circuit describes a Trotterized time evolution operator, we can also reorder anticommuting terms. This is because the product of Pauli gadgets in the circuit was a sum of Pauli strings before trotterization:
\begin{equation}
    U=e^{i\sum_n\theta_nP_n}\approx\left(\prod_ne^{i\frac{\theta_n}{r}P_n} \right)^r,
\end{equation}
where $r$ is the number of trotter steps~\cite{nielsen2010quantum}. This trick changes the trotter error, but that is negligable w.r.t. the error reduction obtained from lowering the gate count in the circuit.
This technique has been used in various other compiling algorithms that use Pauli gadgets in their representation~\cite{huang2024redefininglexicographicalorderingoptimizing,rustiq,li2022paulihedral,cowtan2020generic}.

In addition to Pauli gadgets, another important group of operations used in this work are Clifford gates, which are defined as the set of operations $U$ for which it holds that for every signed Pauli string $P$ the conjugation $UPU^\dagger$ is a signed Pauli string~\cite{nielsen2010quantum}. Because conjugation is a one-to-one process, we know that for every signed Pauli string $P$ there has to be a signed Pauli string $P_0$ such that $P=UP_0U^\dagger$. This property leads to the conjugate transpose $U^\dagger$ of a Clifford $U$ to also be Clifford because $U^\dagger PU=U^\dagger UP_0U^\dagger U=P_0$ is a Pauli string.

The rest of this section goes over: Pauli pushing, which uses Clifford operations to decompose Pauli strings; Clifford tableaus, which can be used to optimize circuits that contain only Clifford operations; and hypergraphs, which can be used to describe connectivity constraints for devices that allow operations on more than two qubits.

\subsection{Pauli Pushing}
It is possible to transform any Pauli gadget $e^{i\theta P}$ into another Pauli gadget by pushing a Clifford operator $O$ through it according to the following rule:
\begin{equation}
    e^{i\theta P}U=Ue^{i\theta U^\dagger PU}.
\end{equation}
This rule can be used to synthesize Pauli gadgets into CNOTs and single qubit gates~\cite{huang2024redefininglexicographicalorderingoptimizing, Cowtan2020, rustiq}. The idea is to push Clifford gates through Pauli gadgets in such a way that we transform them into hardware compatible operations.

In order to push Clifford operators through Pauli gadgets, we need to generate them first. This is done by introducing a Clifford operator $U$ and its inverse $U^\dagger$, which equates to adding the identity operator $I$. We can then push $U$ through the original Pauli gadget. This leads to the following transformation
\begin{equation}
    e^{i\theta P}=e^{i\theta P}UU^\dagger=Ue^{i\theta U^\dagger PU}U^\dagger.
\end{equation}

When choosing which Clifford gates to add, there are two goals: the introduced Clifford gate should be compatible with the target device, and the final Pauli gadget, after all gates have been pushed through, should be device compatible. Because Pauli gadgets can contain arbitrary angles $\theta$ and most devices only allow arbitrary angles on single qubit gates, the goal is usually to transform the gadget into one that depends on only one qubit.

The amount of qubits affected by a gadget $e^{i\theta P}$ can be seen from the amount of non-identity letters in the Pauli string $P$, we call this the \textit{arity} of the Pauli string. The arity of the Pauli gadget is equal to the arity of the Pauli string in its exponent.
The goal therefore is to select a sequence of Clifford gates $U_0\dots U_n$ in such a way that $e^{i\theta U^\dagger_m\dots U^\dagger_0PU_0...U_n}$ has an arity of one. This leaves us with the following transformation
\begin{equation}
    e^{i\theta P}=U_0\dots U_ne^{i\theta U^\dagger_n\dots U^\dagger_0PU_0...U_n}U^\dagger_n\dots U^\dagger_0.
\end{equation}

The resulting circuit, from decomposing one Pauli gadget, contains a Clifford part before and after the arbitrary single qubit rotation. As these are Clifford gates, we can push them through other Pauli gadgets. When decomposing Pauli gadgets in sequential order, we can push the Clifford part at the end through the remaining Pauli gadgets. This results in a large section at the end of the circuit that only contains Clifford gates, which can be stored in a Clifford tableau, which are explained in the next section. The same cannot be done with the Clifford part at the beginning as this would undo the decomposition process.

\subsection{Clifford Tableaus}
Signed Pauli strings can be stored in a linear amount of memory, because each qubit only adds one Pauli letter. As Clifford gates $U$ map signed Pauli strings $U$ to signed Pauli strings $UPU^\dagger$ through conjugation, we can characterize the action of $U$ on $P$ in linear memory by storing $UPU^\dagger$. This can be extended to a sequence of Clifford operations as $U_n\dots U_0PU_0^\dagger\dots U^\dagger_n$ is still a Pauli string, and tells us how the sequence $U_0\dots U_n$ alters $P$.

We can extend this representation to track how Clifford gates act on an arbitrary amount of qubits by tracking two Pauli strings for each qubit. One Pauli string tracks how the Clifford gates alter $Z$ and one that tracks how they alter $X$, these are called the \textit{stabilizer} and \textit{destabilizer}, respectively. This results in a $2n\times2n$ matrix, called the \textit{Clifford Tableau}~\cite{Aaronson2004}. Note that we do not need to track how the Clifford gates act on $Y$ because $ZX=iY=-XZ$.

This means that we can represent the effect of a limitless amount of Clifford gates in a quadratically sized representation. Hence, the Clifford tableau can also be used for circuit optimization as the number of Clifford gates that can be extracted from a Clifford tableau bounded by its size~\cite{winderl2023architecture}. 
This can lead to a smaller output circuit if the number of merged Clifford gates is large enough.

\subsection{Hypergraphs}
In most cases, connectivity constraints are described using normal graphs. Since normal graphs have edges between two nodes, they are well suited to describe connectivity constraints of two-qubit gates. As our work allows multi-qubit gates that interact with more than two qubits, we need to move to hypergraphs, which allow edges that connect more than two nodes. \textit{Hypergraphs} are graphs that consist of \textit{hypernodes}, and \textit{hyperedges} which are sets containing any combination of hypernodes. The expression hypernode is not common as it is no different from a regular node, but we use this term to distinguish between nodes in normal and hypergraphs.

We consider two different kinds of connectivity constraints in this work, linear and square grid. When generalizing these graphs to hypergraphs, we add hypernodes to the existing edges, turning them into hyperedges. For example in the case that we would support ten qubit gates, we would add eight hypernodes to each existing edge. The resulting hypergraph has minimal overlap, meaning that any hyperedge only shares a maximum of one hypernode with any other hyperedge.
\section{Methods}\label{methods}
Our approach closely follows previous research that represents the circuit as collections of unordered Pauli gadgets that are synthesized using Pauli pushing rules while collecting Clifford gates into a Clifford tableau at the end of the circuit (e.g. \cite{rustiq, huang2024redefininglexicographicalorderingoptimizing}). Afterwards, the Clifford tableau is decomposed to reach the final output circuit. In our case, we want to generate  gates that are Pauli gadgets of the form $e^{\pm i\frac{\pi}{4} O}$. This means that we need to generalize the Pauli pushing rule for this gadget and then use it to design an algorithm for circuit synthesis from Pauli gadgets and Clifford tableaux.

We assume, without loss of generality, that all our available multi-qubit operations are Clifford gates and have the same arity. For example, a device supporting a 4-qubit native gate does not support 2-qubit or 3-qubit gates. We do this because the native gate sets of quantum computers are often heavily restricted due to the combinatorial nature of calibration. Moreover, any synthesis algorithm supporting a specific arity multi-qubit gate can be extended to support multiple arities by implementing a heuristic for choosing between them. As this case is more restrictive, the expection is that our algorithm will be applicable for more devices. Note that as a result of this restriction, our algorithm only works for even arity multi-qubit gates because odd arity gadgets are not sufficient for synthesizing arbitary-sized gadgets to single qubit gates.

\subsection{Pushing Clifford-angled Pauli Gadgets}\label{subsec:push}
The push rule for gates of the form $e^{\pm i\frac{\pi}{4}O}$ is known as: 
\begin{equation}
    e^{\pm i\frac{\pi}{4}O}e^{i\theta P}=\begin{cases}
    e^{\pm i\theta OP}e^{\pm i\frac{\pi}{4}O},\text{if $P$ and $O$ anticommute}\\
    e^{i\theta P}e^{\pm i\frac{\pi}{4}O},\text{if $P$ and $O$ commute.}
    \end{cases}
    \label{pushrule}
\end{equation}
where $P$ and $O$ are Pauli strings and $\theta \in \mathbf{R}$.
For completeness ,we provide a proof for this equation in Appendix A.

To use Equation~\ref{pushrule} we need to know if $O$ and $P$ commute or anticommute. For this we can use the property that the different non-identity Pauli matrices $X$, $Y$, and $Z$ anticommute with each other, while a matrix trivially commutes with itself and identity. As $O$ and $P$ are tensor products of Pauli matrices, we have
\begin{equation}
    OP=\bigotimes_jO_jP_j=\bigotimes_js_jP_jO_j=(\Pi_js_j)PO
\end{equation}
where $s_j$ is $-1$ if $O_j$ and $P_j$ anticommute and $1$ otherwise. This means that $O$ and $P$ commute iff an even amount of Pauli matrices in the tensor product commute.

When pushing Pauli gadgets, the anticommuting letters $O_a$ and $P_a$ in $OP$ do not change the arity of the Pauli gadget $e^{\pm i\theta OP}$ because $O_aP_a\neq I$. Conversely, each pair of commuting letters $O_c\neq I$ and $P_c$ the arity of the Pauli gadget either increases or decreases by one. This is because if $P_c=I$ then $O_cP_c=O_c$ and if $O_c=P_c$ then $O_cP_c=I$.

Because our goal is to reduce the arity of the Pauli gadgets, we have to choose $O$ in such a manner that it anticommutes with $P$, meaning that $O$ needs to anticommute with $P$ on an odd number of pairs. If the arity of $O$ is odd, this leaves us with an even number of pairs $O_c\neq I$ and $P_c$ that commute. This means that we do an even number of plus one and minus one changes to the arity of the Pauli gadget we are changing, thus keeping the parity of the arity the same. Thus, odd arity choices of $O$ are unable to reduce even arity Pauli gadgets to single qubit gates. However, this is not the case for even arity choices of $O$ which flip the parity of the arity according to similar logic.  
Hence, our method only works with even arity multi-qubit gates.

Furthermore, since $O$ needs to anticommute with $P$, we can commute on at most $n-1$ qubits $j$ where $O_j\neq I$, where $n$ is the arity of $O$. This means that the largest possible arity change is $n-1$. We can make these maximally sized changes when the arity of $P$ is large, but we should not use these changes to decrease $P$ needlessly to arities under $n$. This is because we need to have arity $n$ in order to reach arity one since we can only anticommute on one letter as we would keep more letters otherwise, and every arity change has to trivially be a removal.


\subsection{Using the Pushing Rule for Compilation}
During synthesis, we use three representations for storing the semantics of quantum gates:
\begin{enumerate}
    \item The Circuit, that represents the already solved part of the problem, initially empty.
    \item The Storage, that contains the non-Clifford Pauli gadgets that are not yet handled.
    \item The Clifford tableau, that represents Clifford gates at the end of the circuit.
\end{enumerate}
Initially, the storage and Clifford tableau are filled with the semantics of a given quantum program. For example, a typical circuit in the Clifford+$T$ gate set is read back to front and each $T$ gate is added as a single qubit Pauli gadget to the Storage and each Clifford gate is pushed through the Storage and collected in the Clifford tableau.

Using these representations, the pushing program works is as follows:
\begin{algorithm}[H]
\begin{algorithmic}[1]
\Loop
\State Move all single qubit gadgets from the Storage to the Circuit.
\If{Storage is not empty}
\State Select a Pauli gadget $P$ from the Storage.
\State Generate the Clifford gates $U_0\dots U_k$ which decompose $P$ to a single qubit gate.
\State Push all the generated Cliffords $U_0\dots U_k$ through all the Pauli gadgets in the Storage.
\State Add the inverses $U_k^\dagger\dots U_0^\dagger$ to the circuit.
\State Add $U_0\dots U_k$ to the Clifford tableau.
\EndIf
\EndLoop
\end{algorithmic}
\end{algorithm}
\label{mainalgo}

\newpage
Note that the generated multi-qubit Clifford gates can be chosen to be of type $e^{\pm i\frac{\pi}{4}O}$ where $O$ might not be the same Pauli string as in the natively supported $e^{\pm i\frac{\pi}{4}O^{native}}$ gate. This is fixed by adding the necessary single qubit Clifford gates $C_0...C_n$ before and after the native multi qubit gate such that for each qubit $j: O_j=C_jO_j^{native}C_j^\dagger$.

As we can choose any Pauli String $O$ with arity $n$ when generating the Clifford gates on line 5 of the algorithm, we can choose every gate in such a manner that it changes the arity of $P$ by an uneven count up to $n-1$. In the cases that the arity $l$ of $P$ is greater than $n$, we need to use at least $ceil((l-n)/(n-1))$ gates to convert $P$ to arity $n$. Since we can only do uneven arity changes, one additional gate is sometimes needed to get to an uneven arity gadget that can then be converted to an $n$-arity gadget. Then one final gate is needed to convert from arity $n$ to one. In the case that the arity is smaller than $n$, we need one or two gates, depending of the original arity, to get to the arity $n$, and one more to arity one.

As we only need the arity $l$ of a given $P$ to figure out how many $n$-arity gates are needed for the decomposition in step 5, we can use the arities of the gadgets in the storage to greedily choose the next Pauli gadget on line 4 of the algorithm. The combined equation for the amount of gates $k$ needed for decomposition is:
\begin{itemize}
    \item if $l>=n$:  $k = \text{ceil}((l-n) / (n-1))+p$ where $p\in \{1,2\}$ is chosen such that $l$ and $k$ have different parities. 
    \item if $l<n$: $k=3$ if $l$ is even, else $k=2$.
\end{itemize}
This means that when the supported multi qubit gate has arity $4$, and $P$ has arity:
\begin{itemize}
    \item $8$, we use $k=3$ gates for the arity changes $8\rightarrow5\rightarrow4\rightarrow1$.
    \item $9$, we use $k=4$ gates for the arity changes $9\rightarrow6\rightarrow3\rightarrow4\rightarrow1$.
    \item $3$, we use $k=2$ gates for the arity changes $3\rightarrow4\rightarrow1$.
    \item $2$, we use $k=3$ gates for the arity changes $2\rightarrow3\rightarrow4\rightarrow1$.
\end{itemize}

After running the algorithm we end up with a circuit, and the Clifford tableau. In case that the algorithm started with an empty Clifford tableau, the Clifford part only contains $n$-qubit Clifford gates and they could be directly added to the circuit. If the algorithm generated a lot of Clifford gates, or the Clifford tableau was not empty, we need to synthesize the Clifford tableau next. This is explained in Section~\ref{subsec:ct}.

\subsection{Connectivity}\label{subsec:con}
Some quantum computing technologies have a predefined restriction on which qubits their multi-qubit gates can act. To ensure that our algorithm is also applicable for those machines, we make a small adaptation of our algorithm such that the multi-qubit Clifford gates that are generated always adhere to connectivity constraints of the target device. 

For two-qubit gates, the connectivity constraints are represented in a coupling graph. But since our native multi-qubit gate acts on more than two qubits, our coupling graph is a hypergraph. From the connectivity hypergraph, we calculate a normal graph that represents the connectivity constraints between the hyperedges. We call this graph the \textit{Explosion} of the hypergraph and it contains two kinds of nodes: the \textit{centroid} nodes that each correspond to one hyperedge and the hypernodes that are only in said hyperedge, and the \textit{non-centroid} nodes that correspond to hypernodes that share the same hyperedges(plural).

\newpage
The Explosion graph is generated as follows:
\begin{algorithm}[H]
\begin{algorithmic}[1]
\State $G\leftarrow$ Empty graph
\State $centroids\leftarrow $ Centroid node for each hyperedge $E$ in the connectivity hypergraph
\State Add $centroids$ to $G$
\Loop ~$\forall (subset\subseteq hyperedges: |subset|>1)$
\State $shared\leftarrow$ Hypernodes belonging, and only belonging, to all elements of $subset$
\If{$shared$ is not empty}
\State $node\leftarrow$ non-centroid node corresponding to $subset$ and $shared$.
\State Add $node$ to $G$
\Loop ~$\forall (hyperedge\in subset)$
\State $centroid\leftarrow$ The centroid corresponding to $hyperedge$
\State Add edges $(node, centroid)$ to $G$
\EndLoop
\EndIf
\EndLoop
\end{algorithmic}
\end{algorithm}
\label{explosion}

After creating an Explosion graph of the target connectivity, we start decomposition of a Pauli gadget $P$ by selecting all of the nodes in the explosion that contain hypernodes which correspond to non-identity letters in $P$. Then, we calculate a Steiner tree over the Explosion graph that contains the selected nodes, and process the tree by iterating over leaf nodes:
\begin{algorithm}[H]
\begin{algorithmic}[1]
\State $S\leftarrow$ Calculated (approximate) Steiner tree
\State $gates\leftarrow$ Empty sequence
\Loop ~While $S$ is not empty
\State $leaf\leftarrow$ Select leaf node from $S$
\If{$leaf$ is a centroid node}
\State $qubit\leftarrow$ any hypernode in $leaf$ that is also in the parent of $leaf$
\State $newGates\leftarrow$ Clifford gate sequence to be pushed, which only acts on the hyperedge of the centroid and transforms $P$ in such a manner that it only depends on $qubit$ inside of the hyperedge
\State Add $newGates$ to $gates$
\EndIf
\State Remove $leaf$ from $S$
\EndLoop
\State \Return $gates$
\end{algorithmic}
\end{algorithm}

\begin{figure}[H]
    \centering
    \includegraphics[width=0.5\linewidth]{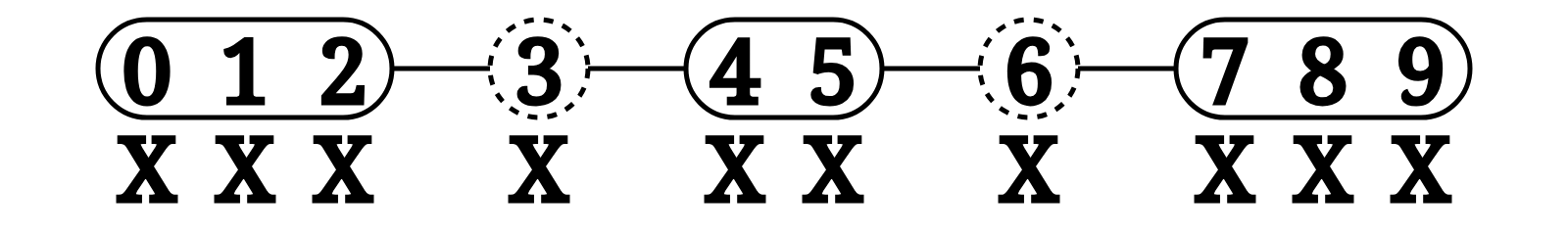}\\
    The leftmost leaf can be solved with $XXXY$ on qubits $0123$.\\
    \includegraphics[width=0.5\linewidth]{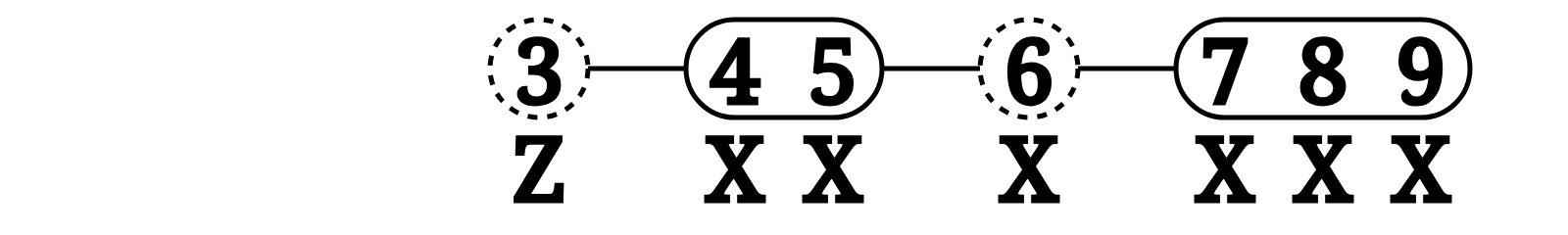}\\
    Similarly the rightmost leaf can be solved with $YXXX$ on qubits $6789$.\\
    \includegraphics[width=0.5\linewidth]{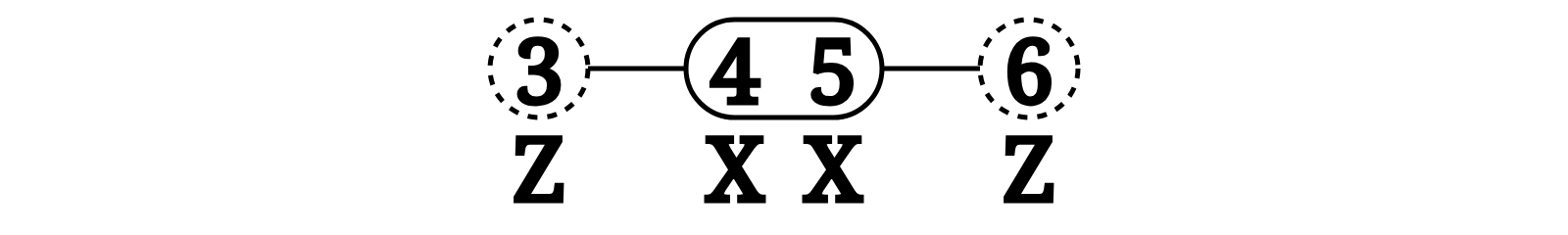}\\
    Now both the leftmost and rightmost leafs are non-centroids and can be removed.\\
    \includegraphics[width=0.5\linewidth]{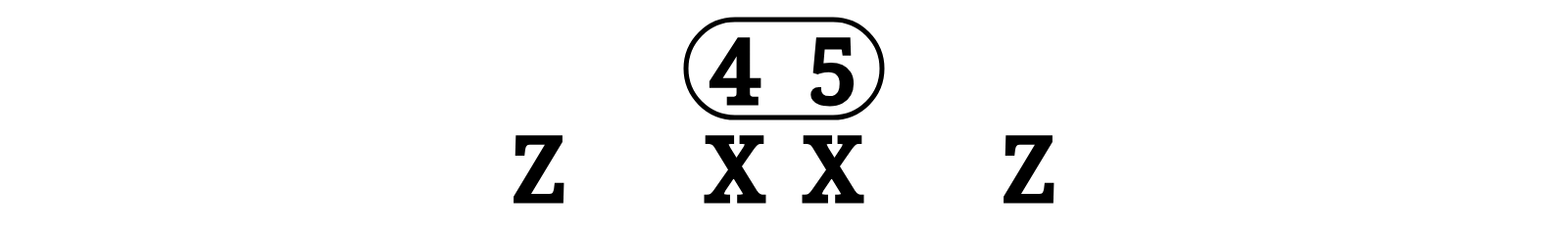}\\
    The last centroid corresponds to qubits $3456$ and we can use $ZYXZ$ to convert the remaining Pauli string to arity one.\\
    \includegraphics[width=0.5\linewidth]{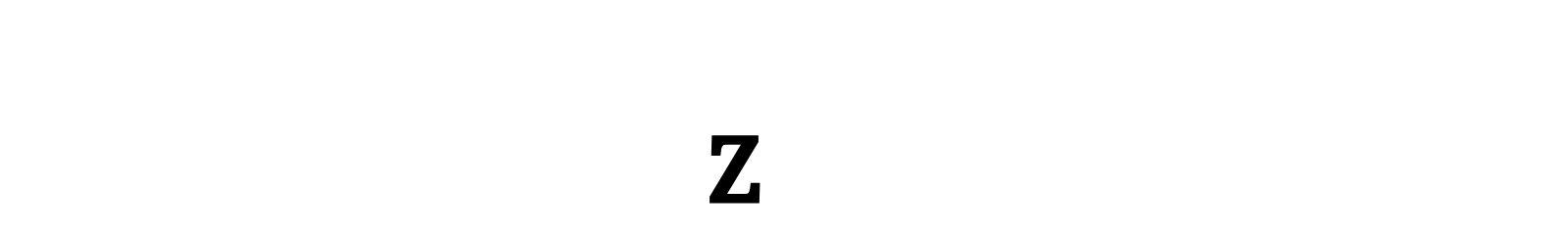}\\
    \caption{A step by step demonstration of how we can decompose a Pauli gadget under the connectivity constraint $\{\{0, 1, 2 ,3\}, \{3, 4, 5, 6\}, \{6, 7, 8, 9\}\}$. The centroids corresponding to the hyperedges are represented as nodes with solid lines, while the non-centroids that contain nodes which are shared between multiple hyperedges are represented with dotted lines. The gadget that is being decomposed consists of $X$-letters on every qubit, which leads to a Steiner tree that contains the complete Explosion graph.}
    \label{fig:connectivityexample}
\end{figure}

This algorithm follows connectivity constraints, because we only use gates on the corresponding hyperedges of centroids. Additionally, we always remove centroids from the problem in such a way that all of the non-identity letters of $P$ can be accessed from remaining centroids, which ultimately ends in a situation where $P$ only depends on one centroid that is used to turn it into a single qubit gadget. We give an example of the decomposition process under a connectivity constraint in Figure~\ref{fig:connectivityexample}.

\subsection{Clifford Tableau Decomposition}\label{subsec:ct}
After synthesis, we are left with a Clifford tableau that we need to decompose back into gates. This is done by iteratively adding a Cliffords gate to the end of the circuit and prepending its inverse to the tableau until the tableau is the identity matrix.

Since adding an operation to the tableau performs the same conjugation (on all of the tableau rows) as Pauli pushing does to the Pauli string of Pauli gadgets, we can utilize the same approach as in Pauli pushing to solve the first row. Firstly, we need to select a row to be solved, this can be done similarly to Pauli gadget decomposition, but now we need to end up with a single qubit operation on a specific qubit. This means that if the specific qubit starts with an identity letter, we need to do additional work to add a non-identity letter to said qubit.

In addition, we need to make sure that the remaining letter corresponds to the original letter $X$ or $Z$ depending on the row, and that the row does not have a negative sign. Since the row that is being handled is solved to only depend on one qubit, this can be achieved using single qubit Clifford operations.

Then, assuming that the solved the row is $G_{i+1}=I^{\otimes i}\otimes  G\otimes I^{\otimes(n-i-1)}$, where $i\in\{0,\dots,n-1\}$, we solve the next row $G^c_{i+1}$ that has to be converted to $I^{\otimes i}\otimes  G^c\otimes I^{\otimes(n-i-1)}$, where $G, G^c\in\{X, Z\}$ and $G^c\neq G$. While doing this, we need to be careful to not alter the already solved row. Our solution is to protect the solved row, by only using the letters $I$ or $G$ on qubit $i+1$, which makes it so that we always commute with the solved row, which keeps it unchanged.

\newpage
Since we need to protect qubit $i+1$, we are more limited in what we are doing, and have to be sure that we can still solve the row $G^c_{i+1}$. Because the rows $G^c_{i+1}$ and $G_{i+1}$ always anticommute~\cite{Aaronson2004}, the row $G^c_{i+1}$ can never have the letter $G$ on qubit $i+1$ as it would otherwise commute with $G_{i+1}$. Due to this property we can always use $G$ to anticommute with the existing letter on $i+1$ to change it between $G^c$ and $Y$. This lets us always choose suitable Clifford gadgets to convert $G^c_{i+1}$ in such a manner that it only depends on qubit $i+1$. Furthermore, the target letter can only be either already solved to be $G^c$, or be $Y$ which we can convert to $G^c$ using a single qubit Clifford gadget with the letter $G$ on qubit $i+1$. A possible negative sign can be removed with two repetitions of a single qubit Clifford gadgets with the letter $G$ on qubit $i+1$.

This leads to the situation where we have solved two rows, which correspond to $X_i$ and $Z_i$, back to their original forms. These rows always anticommute among themselves, but commute with every other row of the Tableau~\cite{Aaronson2004}, which means that all other rows have the identity letter on qubit $i+1$, since otherwise the row would anticommute with one or both of the solved rows. This means that we have completely solved the qubit $i+1$ and we can remove it from the tableau given that we have more than $n$ qubits left. This limitation is in place, because if we have less than $n$ qubits, we cannot apply multi qubit gates on the remaining qubits. We repeat this approach until we are left with only $n-1$ unsolved qubits. 

While we have $2$ to $n-1$ unsolved qubits, we have to change our approach, because we need to use multi-qubit operation which overlap with already solved qubits. This led us to use brute force search to find Clifford gadget sequences that solve rows while keeping already solved rows intact. The found sequences were then generalized to a set of four different sequence types which work for the following four cases:
\begin{itemize}
    \item the target qubit has an even arity string with identity on the target qubit,
    \item the target qubit has an uneven arity string with identity on the target qubit,
    \item the target qubit has an even arity string with non-identity on the target qubit,
    \item the target qubit has an uneven string with non-identity on the target qubit.
\end{itemize}
As these four cases cover all possible situations, we can use this set of sequence to solve the remaining tableau rows. The specific sequences and corresponding proofs can be found Appendix B.

In the case where we have connectivity constraints, we use the explosion of the connectivity graph to create an approximate minimum strainer tree over the whole graph. Which is followed by iteratively removing leafs from said Steiner tree, while solving the unsolved rows that belong to removed centroids. Importantly, each time that we solve a row, we are careful to not disturb already solved rows and to follow the connectivity constraints as explained in the previous section \ref{subsec:con}.

\section{Experiments}\label{experiments}
We ran some experiments to empirically investigate changes in output circuit size, when changing the arity of the supported multi-qubit gate. Additionally, we demonstrate that the size of the Clifford tableau decompositions does not keep increasing when the circuit size is increased. We preface these results with the caveat that our goal was to create a working algorithm, and we leave improvement of the gate count and depth to future work.

We test our algorithm on sequences of randomly generated Pauli gadgets that are assumed to be from trotterization,  which means that we are allowed to reorder terms, even in cases where they don't commute. Gadgets are generated to act on $1$ to $100$ qubits with uniformly sampled placements and uniformly chosen Pauli letters. The angles of the gadgets are uniformly random values between $0$ and $\pi$. Our results are averaged over $100$ different random circuits.

We count the number of multi-qubit gates in the output circuits. This is the metric of interest because currently multi-qubit operations are slower and noisier than single-qubit ones. As we are only interested in the multi-qubit gate count, we only need to specify the arity of the native multi-qubit gate. The exact Pauli string $O^{native}$ is irrelevant as we can transform any multi-qubit gate with the same arity into $O^{native}$ using single qubit Cliffords. Our implementation can be found on Github\footnote{\url{https://github.com/QuantumHel/TEST/tree/b5d28c40ddf25d24bbf4e967a9fa94e3376030ae}}.

The results are shown in 3D plots where we change the arity of the native multi-qubit gate and number of initial Pauli gadgets. We vary the arity between $2$ and $20$ and the number of gadgets between $1$ and $700$.
For the connectivity constraints, we used three common settings: no constraints, linear, and square grid. The results are shown in figures \ref{fig:randomfull}, \ref{fig:randomlinear}, and \ref{fig:randomsquaregrid}, respectively. Each figure consists of three subfigures that represent the gate count from Pauli gadget synthesis (base), the gate count from Clifford tableau synthesis, and the gate count of the total circuit. For completeness, we have added similar graphs for gate depth in Appendix C as well as gate counts for select molecular UCCSD ansatze in Appendix D.

\newcommand\wid{.37}
\newcommand\hsp{-0.8cm}
\newcommand\vsp{-0.5cm}

\begin{figure}[h!]
    \centering
    \begin{subfigure}{\wid\linewidth}
    \hspace*{\hsp}\includegraphics[trim={0.6250cm 0 0 0},clip,width=\linewidth]{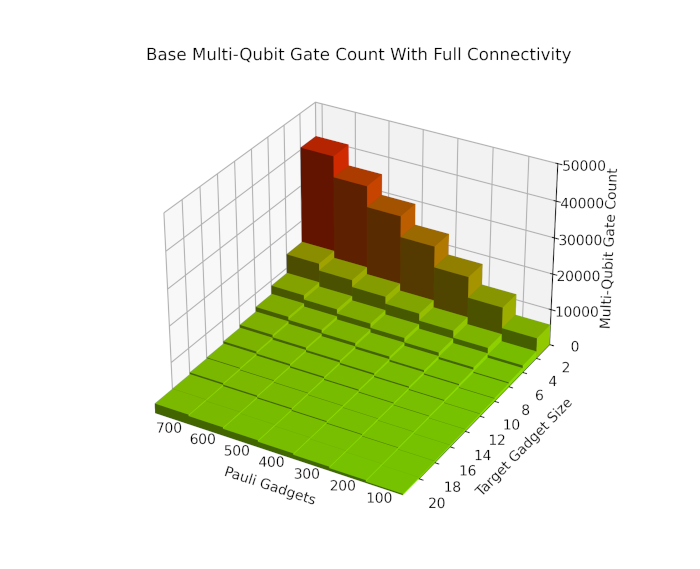}
    \end{subfigure}%
    \begin{subfigure}{\wid\linewidth}
    \hspace*{\hsp}\includegraphics[trim={0.6250cm 0 0 0},clip,width=\linewidth]{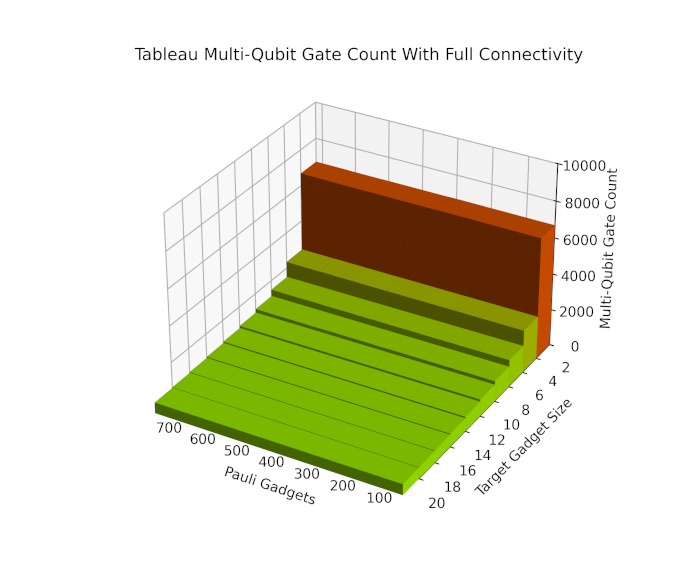}
    \end{subfigure}%
    \begin{subfigure}{\wid\linewidth}
    \hspace*{\hsp}\includegraphics[trim={0.6250cm 0 0 0},clip,width=\linewidth]{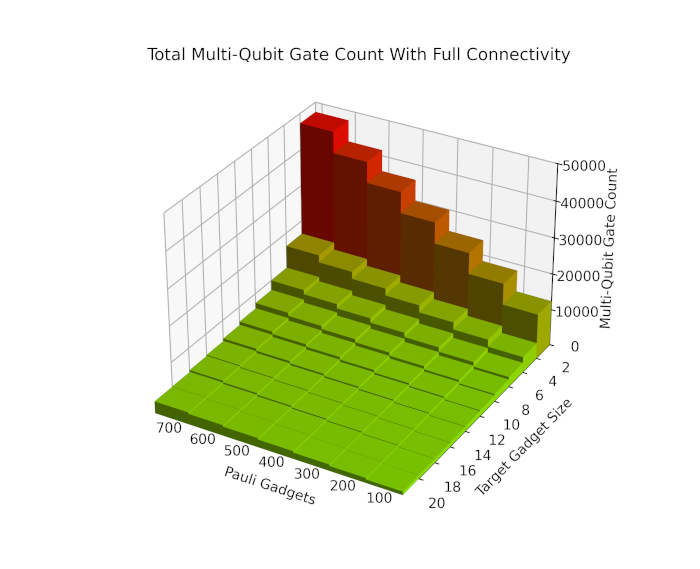}
    \end{subfigure}%
    \caption{The results for the experiment without connectivity constraints.}
    \label{fig:randomfull}
\end{figure}
\vspace{\vsp}
\begin{figure}[h!]
    \centering
    \begin{subfigure}{\wid\linewidth}
    \hspace*{\hsp}\includegraphics[trim={0.6250cm 0 0 0},clip,width=\linewidth]{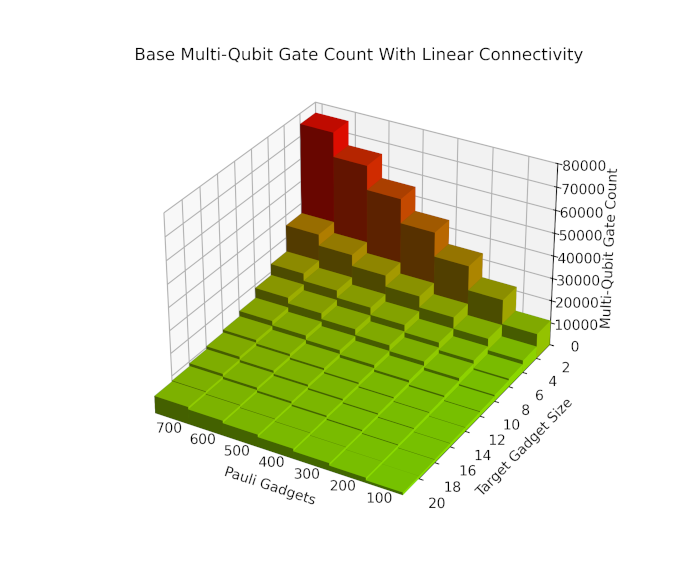}
    \end{subfigure}%
    \begin{subfigure}{\wid\linewidth}
    \hspace*{\hsp}\includegraphics[trim={0.6250cm 0 0 0},clip,width=\linewidth]{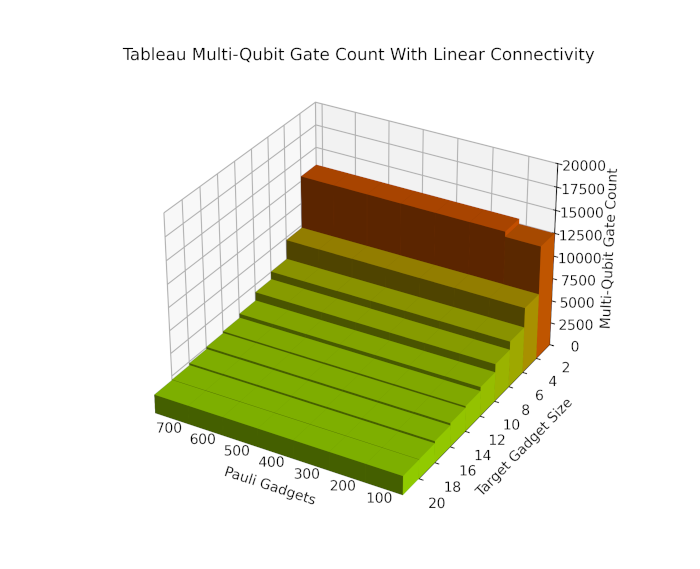}
    \end{subfigure}%
    \begin{subfigure}{\wid\linewidth}
    \hspace*{\hsp}\includegraphics[trim={0.6250cm 0 0 0},clip,width=\linewidth]{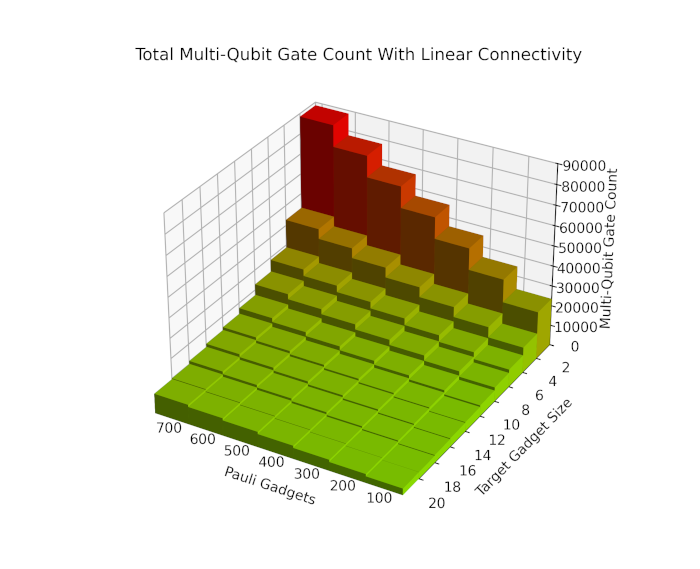}
    \end{subfigure}%
    \caption{The results for the experiment under a linear connectivity constraint.}
    \label{fig:randomlinear}
\end{figure}
\vspace{\vsp}
\begin{figure}[h!]
    \centering
    \begin{subfigure}{\wid\linewidth}
    \hspace*{\hsp}\includegraphics[trim={0.6250cm 0 0 0},clip,width=\linewidth]{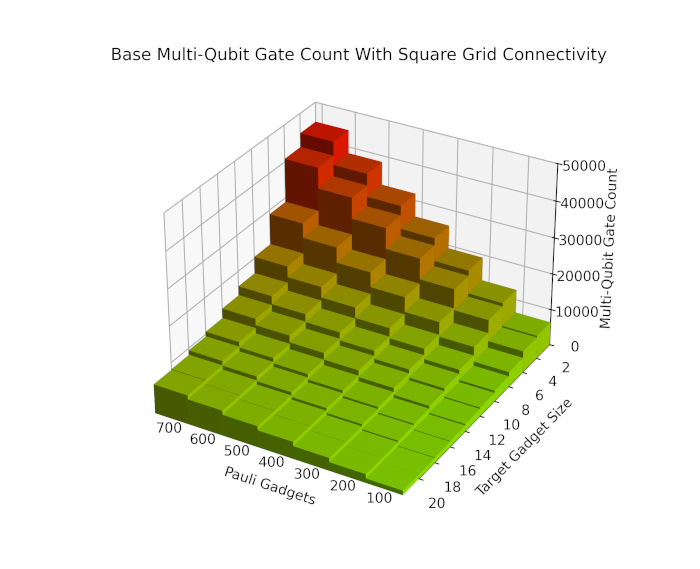}
    \end{subfigure}%
    \begin{subfigure}{\wid\linewidth}
    \hspace*{\hsp}\includegraphics[trim={0.6250cm 0 0 0},clip,width=\linewidth]{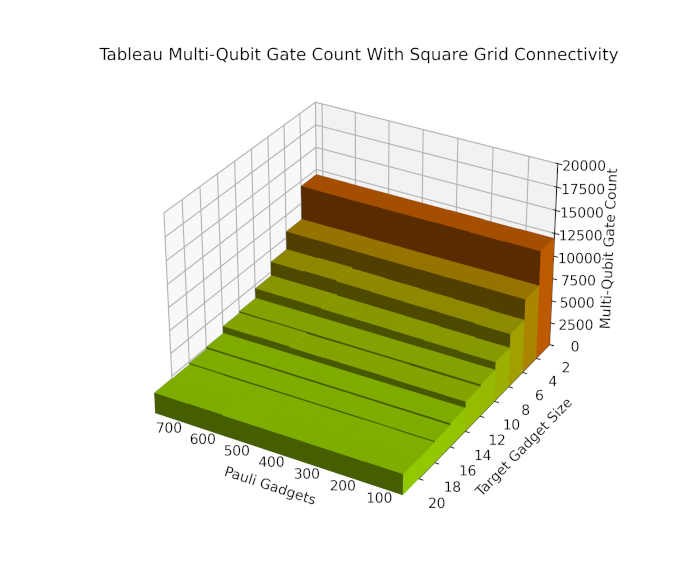}
    \end{subfigure}%
    \begin{subfigure}{\wid\linewidth}
    \hspace*{\hsp}\includegraphics[trim={0.6250cm 0 0 0},clip,width=\linewidth]{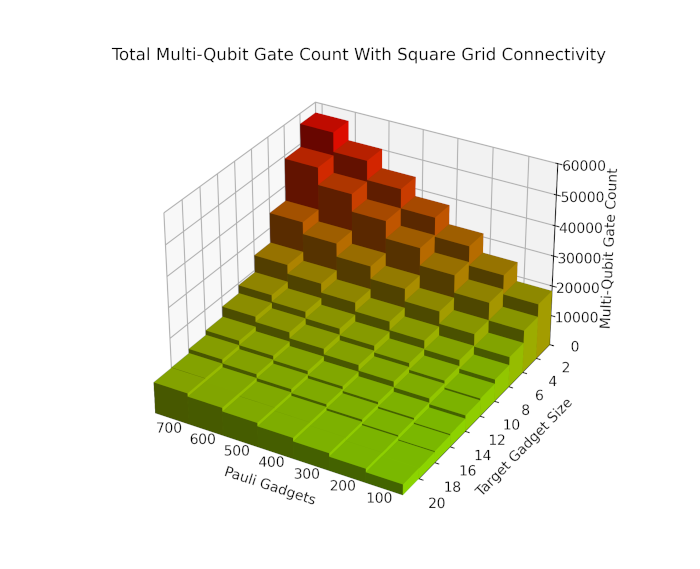}
    \end{subfigure}%
    \caption{The results for the experiment under a square grid connectivity constraint.}
    \label{fig:randomsquaregrid}
\end{figure}

\newpage
Our results indicate that with full and linear connectivity the gate count decreases sharply when moving to 4-qubit gates from 2-qubit gates, but there are diminishing returns when increasing the gate size further. The square grid connectivity differs from this as the decrease from 2- to 4-qubit gates is comparatively small, but the step from 4- to 6-qubit gates is larger, after which we start to see diminishing returns similarly to the previous cases. This indicating that there might be a gate size sweetspot that depends on the used connectivity constraint. Additionally, the plots indicating the size of the decomposed Clifford tableau confirm that the size of the tableau decomposition becomes effectively constant in the input size after a sufficient input size is reached as is consistent with previous work~\cite{patelmarkovhayes, winderl2023architecture}.

\section{Conclusion and Discussion}\label{discussion}
This paper focused on creating an algorithm for circuit synthesis for target gatesets that support multi-qubit Clifford gates of the form $e^{\pm i\frac{\pi}{4} O}$ of arbitrary even arity. In addition, we adapted this algorithm to generate gates that adhere to qubit connectivity constraints. 

The setting for our algorithm is rather artificial in the sense that we assume native non-Cliffords are single qubit gates, single qubit Cliffords can be natively implemented, and the multi-qubit gate has a very particular shape. For example, it is possible that the native entangling gate does not have a single Pauli string, but a sum of Pauli strings, e.g. $O^{native} = XX+YY$. Similarly, a device might support multi-qubit gates with different arities, e.g. $O^{native}$ is a Pauli string of arbitrary size. These cases require some adaptation to our algorithm, which we leave as future work. In principle, the base algorithm works for any natively supported multi-qubit Cliffords, but this generalization requires more complicated rules for pushing and selecting the gates. Our algorithm is just a first step in this direction. 

As it is the first step, the algorithm has not yet been optimized for efficiency in the number of generated gates, e.g. good heuristics for choosing gadget decomposition order. Moreover, it has not been optimized for computational time either. There is much work to be done and we think it is best to target those improvements to specific hardware specifications.

While not optimized, this research can be used to start resource estimation for alternative gatesets that allow higher arity gates. 
Our experiments indicate that supporting higher arity gates can greatly improve circuit size. We leave it as future work to figure out in which circumstances it would be worthwhile for hardware manufacturers to design chips with these alternative gatesets. We hope that our research inspires more general compilation approaches such that we can pin down the best QPU designs.

\section*{Acknowledgements}
This research was funded by Business Finland
project "Enhanced Middleware for Quantum Systems" (EM4QS).

\bibliographystyle{eptcs}
\bibliography{generic}

\newpage
\section*{Appendix A: Proof for Equation~\ref{pushrule}}
\begin{proof}
We start from the property that for each matrix that fulfills $O^2=I$, which is fulfilled by Pauli strings, it holds that
\begin{equation}
    e^{\pm i\frac{\pi}{4}O}=\cos(\frac{\pi}{4})I\pm i\sin(\frac{\pi}{4})O.
\end{equation}
A proof for this can be found in the Picturing Quantum Software book 1.3. \footnote{https://github.com/zxcalc/book/tree/v1.3.0}

This property can be used to transform $\exp(\pm\frac{\pi}{4}O)$ as follows
\begin{equation}
    e^{\pm\frac{\pi}{4}O}
    =\cos(\frac{\pi}{4})I\pm i\sin(\frac{\pi}{4})O
    =\frac{1}{\sqrt{2}}I\pm i\frac{1}{\sqrt{2}}O
    =\frac{1}{\sqrt{2}}(I\pm iO).
\end{equation}
Which gives us
\begin{equation}
    \begin{split}
        e^{\mp i\frac{\pi}{4}O}Pe^{\pm i\frac{\pi}{4}O}
        &=\frac{1}{\sqrt{2}}(I\mp iO)P\frac{1}{\sqrt{2}}(I\pm iO)\\
        &=\frac{1}{2}(P\mp iOP)(I\pm iO)\\
        &=\frac{1}{2}(P\mp iOP\pm iPO+OPO).
    \end{split}
\end{equation}
When $O$ and $P$ commute this simplifies to
\begin{equation}
    \frac{1}{2}(P\mp iOP\pm iPO+OPO)
    =\frac{1}{2}(P\mp iOP\pm iOP+O^2P)
    =\frac{1}{2}(P+IP)
    =P
\end{equation}
and when they anticommute to
\begin{equation}
    \frac{1}{2}(P\mp iOP\pm iPO+OPO)
    =\frac{1}{2}(P\pm iPO\pm iPO-O^2P)
    =\frac{1}{2}(\pm 2iPO)
    =\pm iPO
\end{equation}
As Pauli Matrices always commute or anticommute, Pauli strings inherit this property meaning that the two case above cover all cases.

By applying this to Pauli pushing we get that when $O$ and $P$ commute
\begin{equation}
    e^{\pm i\frac{\pi}{4}O}e^{i\theta P}
    =e^{i\theta \exp(\mp i\frac{\pi}{4}O)P\exp(\pm i\frac{\pi}{4}O)}e^{\pm i\frac{\pi}{4}O}
    =e^{i\theta P}e^{\pm i\frac{\pi}{4}O}
\end{equation}
and when they anticommute
\begin{equation}
    e^{\pm i\frac{\pi}{4}O}e^{i\theta P}
    =e^{i\theta \exp(\mp i\frac{\pi}{4}O)P\exp(\pm i\frac{\pi}{4}O)}e^{\pm i\frac{\pi}{4}O}
    =e^{\pm i\theta OP}e^{\pm i\frac{\pi}{4}O}.
\end{equation}
\end{proof}
\newpage
\section*{Appendix B: Sequences for Decomposing Pauli Tableaus}
When using a multi-qubit gates of size $n$ on $n>k>1$ or less unsolved qubits, we have to operate on already solved qubits that should not be altered. For this we have found four different kinds of gate sequences that cover all  possible cases. We split the cases into those that have an identity gate on the target qubit on its string, and those who have a non identity, and then we further split depending of the evenness of the arity of the original string. We could use a larger amount of more specialized cases to get better performance, but the goal of this paper is to show that this approach works, and adding more cases would lead to an increase in complexity.

These proofs use some special notions:
\begin{itemize}
    \item Subscripts to describe qubit locations. The qubit that we want to solve is $t$ (target), the possibly already solved qubits that we need to operate on that are $I$ in the row that we want to solve are $u$ (uninvolved), and all of the other qubits that are not mentioned to be special cases are $d$ (dirty).
    \item In addition to directly writing the Pauli letters we use the name $O$ to represent the Pauli letter that is on the same index in the original unsolved row that we are trying to solve, the name $G\in\{X, Z\}$ to represent our goal letter and $G^c$ to represent the other letter in $\{X, Z\}$.
    \item We use superscripts to change letters one step in the sequence $X->Z->Y->X$. For example: $X^{++}=Y$ and $O^{+++}=O$.
    \item Since commutation only depends on the evenness of the count of anticommuting letters, the amount of $u$s and $d$s shown depends on the evenness: two for even counts, and one for uneven.
    \item When we sow sequences they are top down sequences in order off pushing, and the proofs have sections with two strings, the upper one is the original string or result from the previous step, and the lower one is the string that we push trough.
\end{itemize}

\subsection*{The Target Qubit has an Even Arity String with Identity on the Target Qubit:}
The sequence consist of 6 strings, where we name one of the dirty qubits as $f$:

\begin{alignat*}{3}
    O_f^+&O_d^+&G_t&Y_u\\
    O_f^{++}&O_d^{++}&G_t^c&X_u\\
    O_f&O_d^{++}&G_t&Z_u\\
    O_f^{++}&O_d&G_t^c&Y_u\\
    O_f^{++}&O_d^{++}&G_t^c&X_u\\
    O_f^+&O_d^+&G_t&Y_u
\end{alignat*}

The left side shows that we solve the row for the qubit that we are solving. The center shows that solved $X_s$ rows stay the same, and the right one that solved $Z_s$ rows stay the same. As the target qubit is identity, we can ignore what happens to the row corresponding to $G^c_t$ as it can not be solved yet. This is because the two rows corresponding to the same letter have to anticommute, and the other row would only have a letter on qubit $t$ with which we commute as $t$ starts from $I$\cite{Aaronson2004} .

\noindent\begin{minipage}{.333\linewidth}
\begin{alignat*}{3}
    O_f&O_d&I_t&I_u\\
    O_f^+&O_d^+&G_t&Y_u\\
    &&&\\
    O_f&O_d&I_t&I_u\\
    O_f^{++}&O_d^{++}&G_t^c&X_u\\
    &&&\\
    O_f&O_d&I_t&I_u\\
    O_f&O_d^{++}&G_t&Z_u\\
    &&&\\
    I_f&O_d^+&G_t&Z_u\\
    O_f^{++}&O_d&G_t^c&Y_u\\
    &&&\\
    O_f^{++}&O_d^{++}&Y&X_u\\
    O_f^{++}&O_d^{++}&G_t^c&X_u\\
    &&&\\
    I_f&I_d&G_t&I_u\\
    O_f^+&O_d^+&G_t&Y_u\\
    &&&\\
    I_f&I_d&G_t&I_u
\end{alignat*}
\end{minipage}
\noindent\begin{minipage}{.333\linewidth}
\begin{alignat*}{5}
    I_f&I_d&I_t&X_s&I_u&I_u\\
    O_f^+&O_d^+&G_t&Y_s&Y_u&Y_u\\
    &&&\\
    O_f^+&O_d^+&G_t&Z_s&Y_u&Y_u\\
    O_f^{++}&O_d^{++}&G_t^c&X_s&X_u&X_u\\
    &&&\\
    O_f^+&O_d^+&G_t&Z_s&Y_u&Y_u\\
    O_f&O_d^{++}&G_t&Z_s&Z_u&Z_u\\
    &&&\\
    O_f^+&O_d^+&G_t&Z_s&Y_u&Y_u\\
    O_f^{++}&O_d&G_t^c&Y_s&Y_u&Y_u\\
    &&&\\
    O_f^+&O_d^+&G_t&Z_s&Y_u&Y_u\\
    O_f^{++}&O_d^{++}&G_t^c&X_s&X_u&X_u\\
    &&&\\
    O_f^+&O_d^+&G_t&Z_s&Y_u&Y_u\\
    O_f^+&O_d^+&G_t&Y_s&Y_u&Y_u\\
    &&&\\
    I_f&I_d&I_t&X_s&I_u&I_u\\
\end{alignat*}
\end{minipage}
\noindent\begin{minipage}{.333\linewidth}
\begin{alignat*}{5}
    I_f&I_d&I_t&Z_s&I_u&I_u\\
    O_f^+&O_d^+&G_t&Y_s&Y_u&Y_u\\
    &&&\\
    O_f^+&O_d^+&G_t&X_s&Y_u&Y_u\\
    O_f^{++}&O_d^{++}&G_t^c&X_s&X_u&X_u\\
    &&&\\
    O_f&O_d&Y_t&I_s&Z_u&Z_u\\
    O_f&O_d^{++}&G_t&Z_s&Z_u&Z_u\\
    &&&\\
    O_f&O_d&Y_t&I_s&Z_u&Z_u\\
    O_f^{++}&O_d&G_t^c&Y_s&Y_u&Y_u\\
    &&&\\
    O_f&O_d&Y_t&I_s&Z_u&Z_u\\
    O_f^{++}&O_d^{++}&G_t^c&X_s&X_u&X_u\\
    &&&\\
    O_f^+&O_d^+&G_t&X_s&Y_u&Y_u\\
    O_f^+&O_d^+&G_t&Y_s&Y_u&Y_u\\
    &&&\\
    I_f&I_d&I_t&Z_s&I_u&I_u\\
\end{alignat*}
\end{minipage}

\subsection*{The Target Qubit has an Uneven Arity String with Identity on the Target Qubit.}
The sequence consist of 6 strings:

\begin{alignat*}{3}
    O^+_d&Y_t&X_u&X_u\\
    O^{++}_d&Y_t&Z_u&Z_u\\
    O^{++}_d&G^c_t&Y_u&Y_u\\
    O^+_d&Y_t&Y_u&Y_u\\
    O^{++}_d&Y_t&Z_u&Z_u\\
    O^+_d&Y_t&X_u&X_u\\
\end{alignat*}

The proof columns are ordered the same as in the previous case:

\noindent\begin{minipage}{.333\linewidth}
\begin{alignat*}{3}
    O_d&I_t&I_u&I_u\\
    O^+_d&Y_t&X_u&X_u\\
    &&&\\
    O^{++}_d&Y_t&X_u&X_u\\
    O^{++}_d&Y_t&Z_u&Z_u\\
    &&&\\
    O^{++}_d&Y_t&X_u&X_u\\
    O^{++}_d&G^c_t&Y_u&Y_u\\
    &&&\\
    I_d&G_t&Z_u&Z_u\\
    O^+_d&Y_t&Y_u&Y_u\\
    &&&\\
    O^+_d&G^c_t&X_u&X_u\\
    O^{++}_d&Y_t&Z_u&Z_u\\
    &&&\\
    O^+_d&G^c_t&X_u&X_u\\
    O^+_d&Y_t&X_u&X_u\\
    &&&\\
    I_d&G_t&I_u&I_u\\
\end{alignat*}
\end{minipage}
\noindent\begin{minipage}{.333\linewidth}
\begin{alignat*}{5}
    I_d&I_t&X_s&I_u\\
    O^+_d&Y_t&X_s&X_u\\
    &&&\\
    I_d&I_t&X_s&I_u\\
    O^{++}_d&Y_t&Z_s&Z_u\\
    &&&\\
    O^{++}_d&Y_t&Y_s&Z_u\\
    O^{++}_d&G^c_t&Y_s&Y_u\\
    &&&\\
    O^{++}_d&Y_t&Y_s&Z_u\\
    O^+_d&Y_t&Y_s&Y_u\\
    &&&\\
    O^{++}_d&Y_t&Y_s&Z_u\\
    O^{++}_d&Y_t&Z_s&Z_u\\
    &&&\\
    I_d&I_t&X_s&I_u\\
    O^+_d&Y_t&X_s&X_u\\
    &&&\\
    I_d&I_t&X_s&I_u\\
\end{alignat*}
\end{minipage}
\noindent\begin{minipage}{.333\linewidth}
\begin{alignat*}{5}
    I_d&I_t&Z_s&I_u\\
    O^+_d&Y_t&X_s&X_u\\
    &&&\\
    O^+_d&Y_t&Y_s&X_u\\
    O^{++}_d&Y_t&Z_s&Z_u\\
    &&&\\
    O_d&I_y&X_s&Y_u\\
    O^{++}_d&G^c_t&Y_s&Y_u\\
    &&&\\
    O_d&I_y&X_s&Y_u\\
    O^+_d&Y_t&Y_s&Y_u\\
    &&&\\
    O_d&I_y&X_s&Y_u\\
    O^{++}_d&Y_t&Z_s&Z_u\\
    &&&\\
    O^+_d&Y_t&Y_s&X_u\\
    O^+_d&Y_t&X_s&X_u\\
    &&&\\
    I_d&I_t&Z_s&I_u\\
\end{alignat*}
\end{minipage}

\subsection*{The Target Qubit has an Even Arity String with Non-identity on the Target Qubit}
If the target qubit already has the correct letter in the correct spot, we use a single qubit gate $G^c$ to change it to another letter. We use the name $A$ to represent the non-identity letter that is not $G$ and not $O_t$(after the possible single qubit gate). The remaining sequence consists of $5$ strings.

\begin{alignat*}{3}
    O_d&G_t&X_u&X_u\\
    O_d&G_t&Z_u&Z_u\\
    O_d&A_t&Y_u&Y_u\\
    O_d&G_t&Z_u&Z_u\\
    O_d&G_t&X_u&X_u\\
\end{alignat*}

We still have on the left that we solve the row target row. Next to it we show that we don't destroy a possibly already solved $G^c_t$ row. In the case that we have solved row $G^c_t$, we know that $O_t$(before the single qubit gate) has to be either $G$ or $Y$, and if it is $G$, we change it to $Y$, meaning that $A$ = $G^c$. The remaining two show that we keep solved $X_s$ and $Z_s$ rows:

\noindent\begin{minipage}{.25\linewidth}
\begin{alignat*}{3}
    O_d&O_t&I_u&I_u\\
    O_d&G_t&X_u&X_u\\
    &&&\\
    I_d&A_t&X_u&X_u\\
    O_d&G_t&Z_u&Z_u\\
    &&&\\
    O_d&O_t&Y_u&Y_u\\
    O_d&A_t&Y_u&Y_u\\
    &&&\\
    I_d&G_t&I_u&I_u\\
    O_d&G_t&Z_u&Z_u\\
    &&&\\
    I_d&G_t&I_u&I_u\\
    O_d&G_t&X_u&X_u\\
    &&&\\
    I_d&G_t&I_u&I_u\\
\end{alignat*}
\end{minipage}
\noindent\begin{minipage}{.25\linewidth}
\begin{alignat*}{5}
    I_d&G^c_t&I_u&I_u\\
    O_d&G_t&X_u&X_u\\
    &&&\\
    O_d&Y_t&X_u&X_u\\
    O_d&G_t&Z_u&Z_u\\
    &&&\\
    I_d&G^c&Y_u&Y_u\\
    O_d&G^c&Y_u&Y_u\\
    &&&\\
    I_d&G^c&Y_u&Y_u\\
    O_d&G_t&Z_u&Z_u\\
    &&&\\
    O_d&Y_t&X_u&X_u\\
    O_d&G_t&X_u&X_u\\
    &&&\\
    I_d&G^c_t&I_u&I_u\\
\end{alignat*}
\end{minipage}
\noindent\begin{minipage}{.25\linewidth}
\begin{alignat*}{5}
    I_d&I_t&X_s&I_u\\
    O_d&G_t&X_s&X_u\\
    &&&\\
    I_d&I_t&X_s&I_u\\
    O_d&G_t&Z_s&Z_u\\
    &&&\\
    O_d&G_t&Y_s&Z_u\\
    O_d&A_t&Y_s&Y_u\\
    &&&\\
    O_d&G_t&Y_s&Z_u\\
    O_d&G_t&Z_s&Z_u\\
    &&&\\
    I_d&I_t&X_s&I_u\\
    O_d&G_t&X_s&X_u\\
    &&&\\
    I_d&I_t&X_s&I_u\\
\end{alignat*}
\end{minipage}
\noindent\begin{minipage}{.25\linewidth}
\begin{alignat*}{5}
    I_d&I_t&Z_s&I_u\\
    O_d&G_t&X_s&X_u\\
    &&&\\
    O_d&G_t&Y_s&X_u\\
    O_d&G_t&Z_s&Z_u\\
    &&&\\
    O_d&G_t&Y_s&X_u\\
    O_d&A_t&Y_s&Y_u\\
    &&&\\
    O_d&G_t&Y_s&X_u\\
    O_d&G_t&Z_s&Z_u\\
    &&&\\
    O_d&G_t&Y_s&X_u\\
    O_d&G_t&X_s&X_u\\
    &&&\\
    I_d&I_t&Z_s&I_u
\end{alignat*}
\end{minipage}

\subsection*{The Target Qubit has an Uneven String with Non-identity on the Target Qubit.}
If the target qubit already has the correct letter in the correct spot, we again use a single qubit gate $G^c$ to change it to another letter, and this result is $O_t$. The remaining sequence consists of $3$ strings:

\begin{alignat*}{3}
    O^+_d&O^+_d&G_t&Y_u\\
    O^{++}_d&O^{++}_d&O_t&Y_u\\
    O^+_d&O^+_d&G_t&Y_u.\\
\end{alignat*}

The proof rows are the same as previously. In the case that we already have a solved row for $G^c_t$, we have originally either $G$ or $Y$ on qubit $t$ and in the case that it is $G$ we change it to $Y$, meaning that $O_t=Y$. We also use the definition of $A$ from the previous sequence.

\noindent\begin{minipage}{.25\linewidth}
\begin{alignat*}{3}
    O_d&O_d&O_t&I_u\\
    O^+_d&O^+_d&G_t&Y_u\\
    &&&\\
    O^{++}_d&O^{++}_d&A_t&Y_u\\
    O^{++}_d&O^{++}_d&O_t&Y_u\\
    &&&\\
    I_d&I_d&G_t&I_u\\
    O^+_d&O^+_d&G_t&Y_u\\
    &&&\\
    I_d&I_d&G_t&I_u\\
\end{alignat*}
\end{minipage}
\noindent\begin{minipage}{.25\linewidth}
\begin{alignat*}{5}
    I_d&I_d&G^c_t&I_u\\
    O^+_d&O^+_d&G_t&Y_u\\
    &&&\\
    O^+_d&O^+_d&Y_t&Y_u\\
    O^{++}_d&O^{++}_d&Y_t&Y_u\\
    &&&\\
    O^+_d&O^+_d&Y_t&Y_u\\
    O^+_d&O^+_d&G_t&Y_u\\
    &&&\\
    I_d&I_d&G^c_t&I_u\\
\end{alignat*}
\end{minipage}
\noindent\begin{minipage}{.25\linewidth}
\begin{alignat*}{5}
    I_d&I_d&I_t&X_s&I_u&I_u\\
    O^+_d&O^+_d&G_t&Y_s&Y_u&Y_u\\
    &&&\\
    O^+_d&O^+_d&G_t&Z_s&Y_u&Y_u\\
    O^{++}_d&O^{++}_d&O_t&Y_s&Y_u&Y_u\\
    &&&\\
    O^+_d&O^+_d&G_t&Z_s&Y_u&Y_u\\
    O^+_d&O^+_d&G_t&Y_s&Y_u&Y_u\\
    &&&\\
    I_d&I_d&I_t&X_s&I_u&I_u\\
\end{alignat*}
\end{minipage}
\noindent\begin{minipage}{.25\linewidth}
\begin{alignat*}{5}
    I_d&I_d&I_t&Z_s&I_u&I_u\\
    O^+_d&O^+_d&G_t&Y_s&Y_u&Y_u\\
    &&&\\
    O^+_d&O^+_d&G_t&X_s&Y_u&Y_u\\
    O^{++}_d&O^{++}_d&O_t&Y_s&Y_u&Y_u\\
    &&&\\
    O^+_d&O^+_d&G_t&X_s&Y_u&Y_u\\
    O^+_d&O^+_d&G_t&Y_s&Y_u&Y_u\\
    &&&\\
    I_d&I_d&I_t&Z_s&I_u&I_u\\
\end{alignat*}
\end{minipage}

\newpage
\section*{Appendix D: Plots for the Depth Behavior of the Randomized Experiments}
\begin{figure}[h]
    \centering
    \begin{subfigure}{.37\linewidth}
    \hspace*{0cm}\includegraphics[trim={0.6250cm 0 0 0},clip,width=\linewidth]{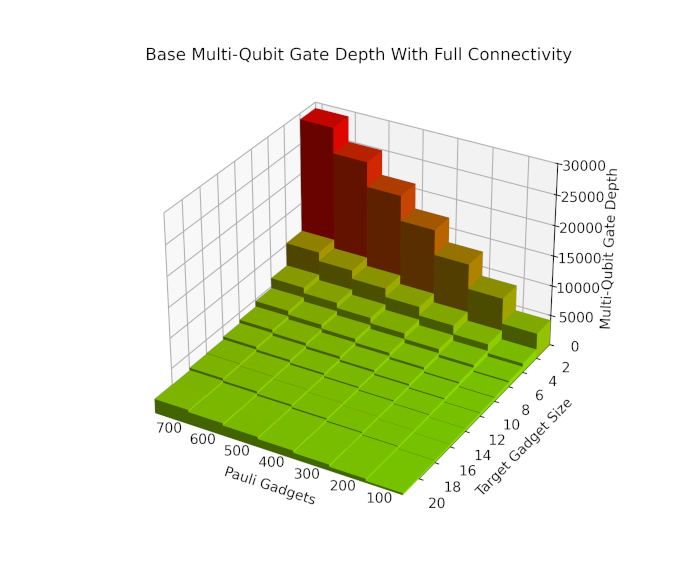}
    \end{subfigure}%
    \begin{subfigure}{.37\linewidth}
    \hspace*{0cm}\includegraphics[trim={0.6250cm 0 0 0},clip,width=\linewidth]{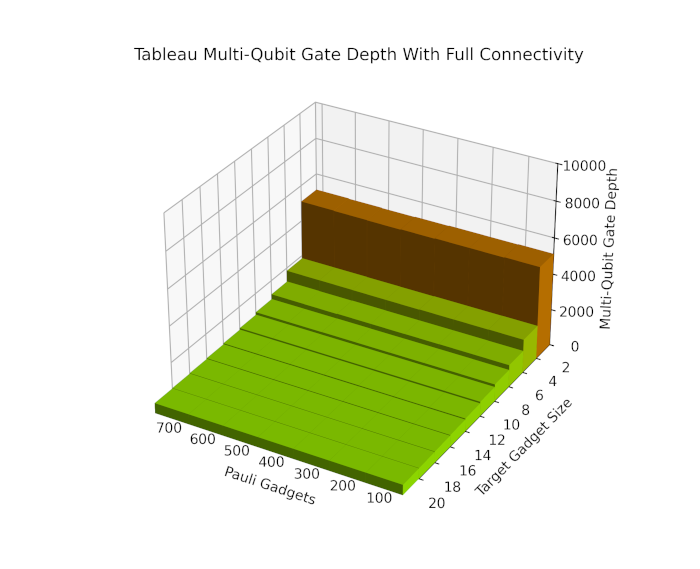}
    \end{subfigure}%
    \begin{subfigure}{.37\linewidth}
    \hspace*{0cm}\includegraphics[trim={0.6250cm 0 0 0},clip,width=\linewidth]{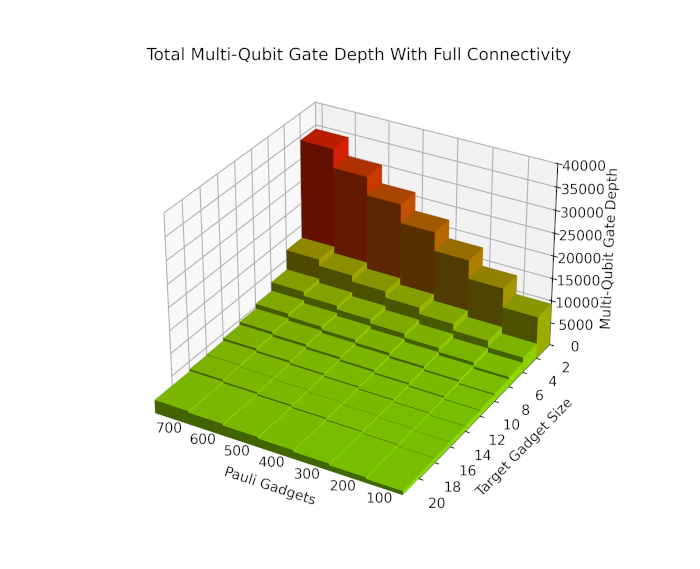}
    \end{subfigure}%
    \caption{The results for the experiment with no connectivity constraints.}
\end{figure}

\begin{figure}[h]
    \centering
    \begin{subfigure}{.37\linewidth}
    \hspace*{0cm}\includegraphics[trim={0.6250cm 0 0 0},clip,width=\linewidth]{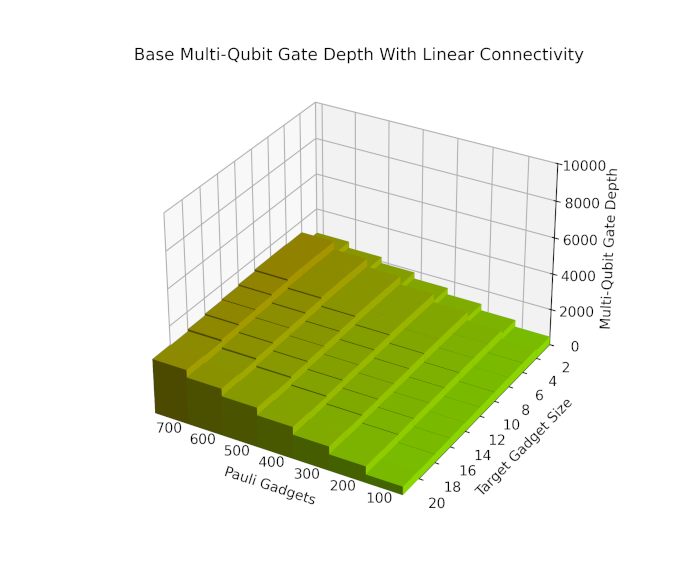}
    \end{subfigure}%
    \begin{subfigure}{.37\linewidth}
    \hspace*{0cm}\includegraphics[trim={0.6250cm 0 0 0},clip,width=\linewidth]{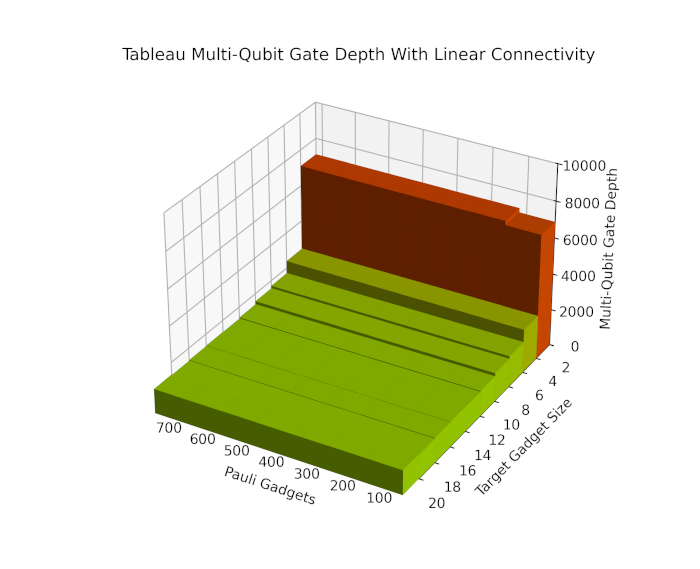}
    \end{subfigure}%
    \begin{subfigure}{.37\linewidth}
    \hspace*{0cm}\includegraphics[trim={0.6250cm 0 0 0},clip,width=\linewidth]{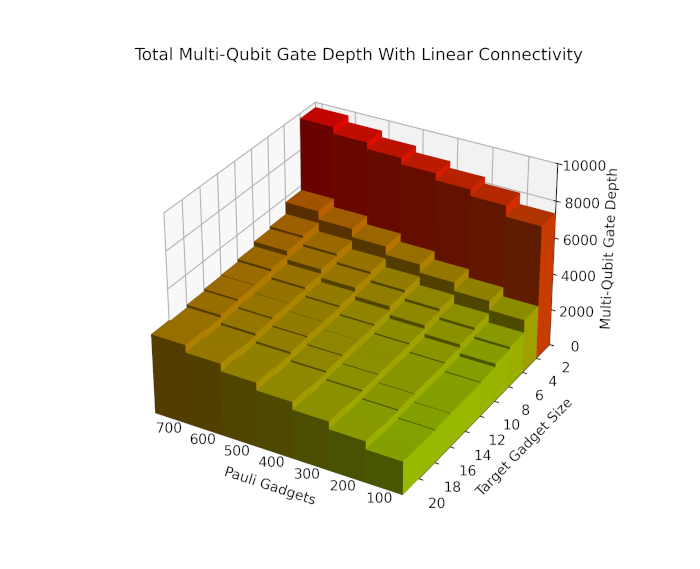}
    \end{subfigure}%
    \caption{The results for the experiment under a linear connectivity constraint.}
\end{figure}

\begin{figure}[hbt!]
    \centering
    \begin{subfigure}{.37\linewidth}
    \hspace*{0cm}\includegraphics[trim={0.6250cm 0 0 0},clip,width=\linewidth]{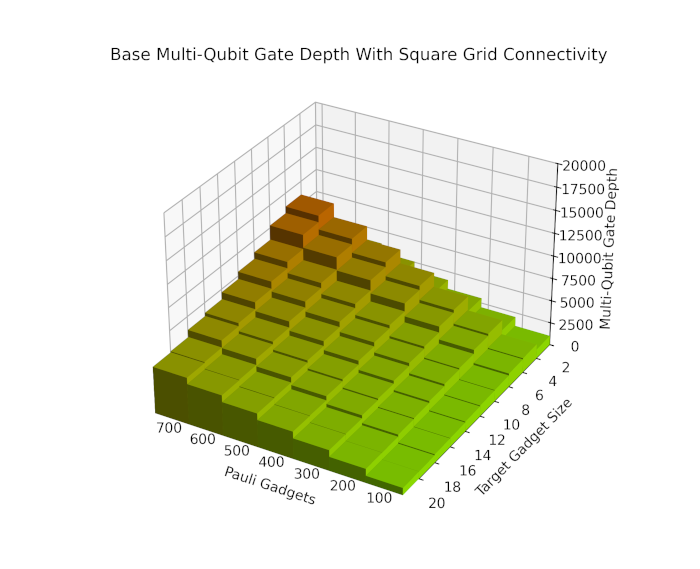}
    \end{subfigure}%
    \begin{subfigure}{.37\linewidth}
    \hspace*{0cm}\includegraphics[trim={0.6250cm 0 0 0},clip,width=\linewidth]{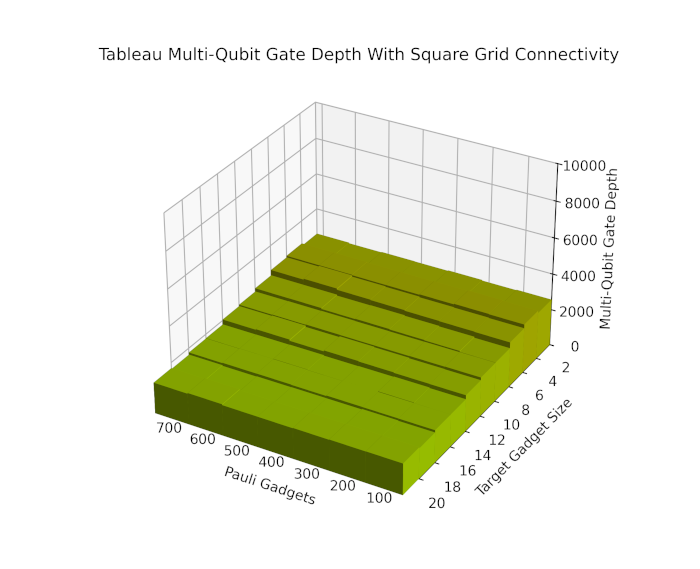}
    \end{subfigure}%
    \begin{subfigure}{.37\linewidth}
    \hspace*{0cm}\includegraphics[trim={0.6250cm 0 0 0},clip,width=\linewidth]{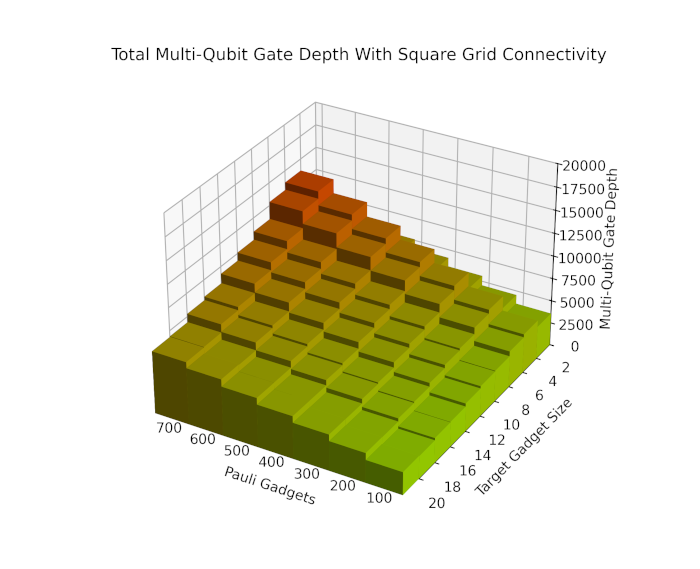}
    \end{subfigure}%
    \caption{The results for the experiment under a square grid connectivity constraint.}
\end{figure}
\newpage
\section*{Appendix D: Additional Experiments on Molecule Data}

For a more realistic dataset that can be compared with other papers, we also run our experiment on a list of molecule data that is identical to that used in the work that our paper builds on\cite{huang2024redefininglexicographicalorderingoptimizing}. The results for gate sizes 2, 4, 6, 10, and 20 are shown in three tables \ref{table:fullconnmolecules}, \ref{table:linearconnmolecules}, and \ref{table:squareconnmolecules} which correspond to full, linear and square  grid connectivities respectively. These results are from single runs, and as the algorithm has some randomness from calculating approximate Steiner trees, these results may vary.

\begin{table}[hbt!]
    \begin{changemargin}{-3cm}{-3cm}
    \centering
    \begin{tabular}{|l|c|g|c|g|c|g|c|g|c|g|c|g|c|g|c|}
        
        \hline
        \multirow{2}{3em}{\textbf{Molecule}} & \multirow{2}{2.8em}{\textbf{Qubits}} & \multicolumn{2}{|c|}{\textbf{Multi-Q Inputs}} & \multicolumn{2}{|c|}{\textbf{2-Q Gates}} & \multicolumn{2}{|c|}{\textbf{4-Q Gates}} & \multicolumn{2}{|c|}{\textbf{6-Q gates}} & \multicolumn{2}{|c|}{\textbf{10-Q Gates}} & \multicolumn{2}{|c|}{\textbf{20-Q Gates}} \\
        \cline{3-14}
        & & count & depth & count & depth & count & depth & count & depth & count & depth & count & depth\\
        \hline
        H2\_P\_sto3g & 2 & 2 & 2 & 4 & 4 & 23 & 23 & 23 & 23 & 23 & 23 & 23 & 23\\\hline
H2\_BK\_sto3g & 4 & 12 & 11 & 24 & 21 & 32 & 32 & 39 & 39 & 39 & 39 & 39 & 39\\\hline
H2\_JW\_sto3g & 4 & 12 & 12 & 25 & 23 & 29 & 29 & 46 & 46 & 46 & 46 & 46 & 46\\\hline
H2\_P\_631g & 6 & 152 & 146 & 190 & 156 & 143 & 143 & 168 & 168 & 184 & 184 & 184 & 184\\\hline
H4\_P\_sto3g & 6 & 158 & 152 & 224 & 182 & 145 & 145 & 167 & 167 & 194 & 194 & 194 & 194\\\hline
H2\_JW\_631g & 8 & 84 & 84 & 213 & 173 & 122 & 121 & 114 & 114 & 164 & 164 & 164 & 164\\\hline
H2\_BK\_631g & 8 & 84 & 83 & 211 & 156 & 117 & 116 & 120 & 120 & 145 & 145 & 145 & 145\\\hline
H4\_BK\_sto3g & 8 & 160 & 157 & 355 & 277 & 187 & 186 & 186 & 186 & 229 & 229 & 229 & 229\\\hline
H4\_JW\_sto3g & 8 & 160 & 160 & 383 & 296 & 200 & 195 & 172 & 172 & 245 & 245 & 245 & 245\\\hline
NH\_BK\_sto3g & 12 & 640 & 637 & 2369 & 1587 & 1047 & 926 & 709 & 707 & 688 & 688 & 797 & 797\\\hline
LiH\_BK\_sto3g & 12 & 640 & 633 & 2313 & 1552 & 1024 & 932 & 714 & 714 & 704 & 704 & 823 & 823\\\hline
LiH\_JW\_sto3g & 12 & 640 & 640 & 2231 & 1485 & 1006 & 900 & 707 & 707 & 685 & 685 & 828 & 828\\\hline
NH\_JW\_sto3g & 12 & 640 & 640 & 2326 & 1512 & 1013 & 922 & 705 & 703 & 682 & 682 & 797 & 797\\\hline
H2O\_P\_sto3g & 12 & 1071 & 1051 & 2828 & 1785 & 1436 & 1319 & 1102 & 1100 & 1062 & 1062 & 1227 & 1227\\\hline
BeH2\_P\_sto3g & 12 & 1075 & 1054 & 2973 & 1954 & 1468 & 1314 & 1108 & 1105 & 1055 & 1055 & 1207 & 1207\\\hline
CH2\_P\_sto3g & 12 & 2093 & 2054 & 5101 & 3189 & 2469 & 2244 & 1992 & 1989 & 1937 & 1937 & 2192 & 2192\\\hline
H2O\_JW\_sto3g & 14 & 1000 & 1000 & 4382 & 2724 & 1962 & 1629 & 1235 & 1228 & 1018 & 1018 & 1226 & 1226\\\hline
H2O\_BK\_sto3g & 14 & 1000 & 995 & 4483 & 2853 & 1944 & 1625 & 1186 & 1180 & 1040 & 1040 & 1219 & 1219\\\hline
CH2\_BK\_sto3g & 14 & 1488 & 1479 & 6196 & 3802 & 2778 & 2366 & 1690 & 1684 & 1517 & 1517 & 1699 & 1699\\\hline
BeH2\_JW\_sto3g & 14 & 1488 & 1488 & 5955 & 3545 & 2698 & 2263 & 1673 & 1651 & 1488 & 1488 & 1710 & 1710\\\hline
BeH2\_BK\_sto3g & 14 & 1488 & 1466 & 6265 & 3874 & 2813 & 2364 & 1700 & 1679 & 1498 & 1498 & 1706 & 1706\\\hline
CH2\_JW\_sto3g & 14 & 1488 & 1488 & 5989 & 3526 & 2736 & 2271 & 1693 & 1677 & 1491 & 1491 & 1713 & 1713\\\hline
H4\_P\_631g & 14 & 2902 & 2868 & 9402 & 5520 & 4539 & 3856 & 2960 & 2943 & 2711 & 2711 & 3091 & 3091\\\hline
H4\_JW\_631g & 16 & 1440 & 1440 & 7428 & 4624 & 2991 & 2397 & 2156 & 2075 & 1478 & 1478 & 1784 & 1784\\\hline
H4\_BK\_631g & 16 & 1440 & 1429 & 7415 & 4698 & 3009 & 2441 & 2054 & 1993 & 1497 & 1497 & 1745 & 1745\\\hline
NH3\_JW\_sto3g & 16 & 2340 & 2340 & 11399 & 6693 & 4779 & 3824 & 3077 & 2988 & 2351 & 2351 & 2680 & 2680\\\hline
NH3\_BK\_sto3g & 16 & 2340 & 2329 & 11102 & 6628 & 4639 & 3737 & 2889 & 2819 & 2369 & 2369 & 2643 & 2643\\\hline
HCl\_JW\_sto3g & 20 & 684 & 684 & 5186 & 3120 & 2113 & 1608 & 1426 & 1247 & 864 & 864 & 1109 & 1109\\\hline
HCl\_BK\_sto3g & 20 & 684 & 682 & 4614 & 2861 & 2076 & 1595 & 1400 & 1227 & 896 & 896 & 1140 & 1140\\\hline    
    \end{tabular}
    \end{changemargin}
    \caption{Full connectivity table for molecular data experiment.}
    \label{table:fullconnmolecules}
\end{table}

\begin{table}[p]
    \begin{changemargin}{-3cm}{-3cm}
    \centering
    \begin{tabular}{|l|c|g|c|g|c|g|c|g|c|g|c|g|c|g|c|}
        
        \hline
        \multirow{2}{3em}{\textbf{Molecule}} & \multirow{2}{2.8em}{\textbf{Qubits}} & \multicolumn{2}{|c|}{\textbf{Multi-Q Inputs}} & \multicolumn{2}{|c|}{\textbf{2-Q Gates}} & \multicolumn{2}{|c|}{\textbf{4-Q Gates}} & \multicolumn{2}{|c|}{\textbf{6-Q gates}} & \multicolumn{2}{|c|}{\textbf{10-Q Gates}} & \multicolumn{2}{|c|}{\textbf{20-Q Gates}} \\
        \cline{3-14}
        & & count & depth & count & depth & count & depth & count & depth & count & depth & count & depth\\
        \hline
        H2\_P\_sto3g & 2 & 2 & 2 & 6 & 6 & 19 & 19 & 22 & 22 & 22 & 22 & 22 & 22\\\hline
H2\_BK\_sto3g & 4 & 12 & 11 & 46 & 42 & 32 & 32 & 55 & 55 & 72 & 72 & 72 & 72\\\hline
H2\_JW\_sto3g & 4 & 12 & 12 & 57 & 50 & 46 & 46 & 56 & 56 & 81 & 81 & 111 & 111\\\hline
H2\_P\_631g & 6 & 152 & 146 & 423 & 279 & 312 & 312 & 199 & 199 & 223 & 223 & 223 & 223\\\hline
H4\_P\_sto3g & 6 & 158 & 152 & 418 & 274 & 316 & 316 & 210 & 210 & 229 & 229 & 229 & 229\\\hline
H2\_JW\_631g & 8 & 84 & 84 & 403 & 267 & 337 & 306 & 272 & 272 & 185 & 185 & 195 & 195\\\hline
H2\_BK\_631g & 8 & 84 & 83 & 424 & 265 & 260 & 243 & 271 & 271 & 176 & 176 & 181 & 181\\\hline
H4\_BK\_sto3g & 8 & 160 & 157 & 679 & 401 & 431 & 404 & 363 & 363 & 243 & 243 & 256 & 256\\\hline
H4\_JW\_sto3g & 8 & 160 & 160 & 723 & 435 & 478 & 427 & 395 & 395 & 262 & 262 & 273 & 273\\\hline
NH\_BK\_sto3g & 12 & 640 & 637 & 4557 & 2102 & 2376 & 1574 & 1574 & 1488 & 1258 & 1258 & 863 & 863\\\hline
LiH\_BK\_sto3g & 12 & 640 & 633 & 4549 & 2129 & 2169 & 1544 & 1552 & 1478 & 1111 & 1111 & 811 & 811\\\hline
LiH\_JW\_sto3g & 12 & 640 & 640 & 4210 & 2038 & 2196 & 1536 & 1592 & 1490 & 1165 & 1165 & 844 & 844\\\hline
NH\_JW\_sto3g & 12 & 640 & 640 & 4453 & 2096 & 2368 & 1600 & 1703 & 1590 & 1283 & 1283 & 860 & 860\\\hline
H2O\_P\_sto3g & 12 & 1071 & 1051 & 6740 & 3185 & 3773 & 2489 & 2484 & 2334 & 1751 & 1751 & 1228 & 1228\\\hline
BeH2\_P\_sto3g & 12 & 1075 & 1054 & 6636 & 3129 & 3822 & 2512 & 2521 & 2367 & 1732 & 1732 & 1260 & 1260\\\hline
CH2\_P\_sto3g & 12 & 2093 & 2054 & 11327 & 5401 & 6796 & 4616 & 4493 & 4267 & 3375 & 3375 & 2241 & 2241\\\hline
H2O\_JW\_sto3g & 14 & 1000 & 1000 & 8271 & 3456 & 4456 & 2647 & 3143 & 2315 & 2251 & 2251 & 1242 & 1242\\\hline
H2O\_BK\_sto3g & 14 & 1000 & 995 & 8623 & 3622 & 4185 & 2532 & 3158 & 2357 & 2191 & 2191 & 1254 & 1254\\\hline
CH2\_BK\_sto3g & 14 & 1488 & 1479 & 12744 & 5481 & 6049 & 3708 & 4159 & 3306 & 3078 & 3078 & 1752 & 1752\\\hline
BeH2\_JW\_sto3g & 14 & 1488 & 1488 & 12829 & 5362 & 5854 & 3694 & 4050 & 3282 & 2880 & 2880 & 1757 & 1757\\\hline
BeH2\_BK\_sto3g & 14 & 1488 & 1466 & 12845 & 5467 & 5638 & 3555 & 4049 & 3296 & 2865 & 2865 & 1695 & 1695\\\hline
CH2\_JW\_sto3g & 14 & 1488 & 1488 & 12608 & 5395 & 6405 & 3827 & 4287 & 3329 & 3087 & 3087 & 1776 & 1776\\\hline
H4\_P\_631g & 14 & 2902 & 2868 & 21123 & 9402 & 10299 & 6613 & 7838 & 5968 & 5282 & 5282 & 3124 & 3124\\\hline
H4\_JW\_631g & 16 & 1440 & 1440 & 14706 & 5535 & 6412 & 3653 & 4283 & 3212 & 2907 & 2907 & 1829 & 1829\\\hline
H4\_BK\_631g & 16 & 1440 & 1429 & 14507 & 5517 & 5943 & 3557 & 4073 & 3190 & 2925 & 2925 & 1759 & 1759\\\hline
NH3\_JW\_sto3g & 16 & 2340 & 2340 & 22963 & 8741 & 10745 & 5886 & 6850 & 5023 & 5065 & 5065 & 2692 & 2692\\\hline
NH3\_BK\_sto3g & 16 & 2340 & 2329 & 23243 & 9165 & 10667 & 5930 & 7086 & 5098 & 4965 & 4965 & 2661 & 2661\\\hline
HCl\_JW\_sto3g & 20 & 684 & 684 & 7175 & 2622 & 5230 & 2315 & 3296 & 2040 & 2434 & 2218 & 1159 & 1159\\\hline
HCl\_BK\_sto3g & 20 & 684 & 682 & 5763 & 2376 & 5198 & 2348 & 3284 & 2091 & 2362 & 2160 & 1156 & 1156\\\hline
    \end{tabular}
    \end{changemargin}
    \caption{Linear connectivity table for molecular data experiment.}
    \label{table:linearconnmolecules}
\end{table}

\begin{table}[p]
    \begin{changemargin}{-3cm}{-3cm}
    \centering
    \begin{tabular}{|l|c|g|c|g|c|g|c|g|c|g|c|g|c|g|c|}
        
        \hline
        \multirow{2}{3em}{\textbf{Molecule}} & \multirow{2}{2.8em}{\textbf{Qubits}} & \multicolumn{2}{|c|}{\textbf{Multi-Q Inputs}} & \multicolumn{2}{|c|}{\textbf{2-Q Gates}} & \multicolumn{2}{|c|}{\textbf{4-Q Gates}} & \multicolumn{2}{|c|}{\textbf{6-Q gates}} & \multicolumn{2}{|c|}{\textbf{10-Q Gates}} & \multicolumn{2}{|c|}{\textbf{20-Q Gates}} \\
        \cline{3-14}
        & & count & depth & count & depth & count & depth & count & depth & count & depth & count & depth\\
        \hline
H2\_P\_sto3g & 2 & 2 & 2 & 6 & 6 & 19 & 19 & 22 & 22 & 22 & 22 & 22 & 22\\\hline
H2\_BK\_sto3g & 4 & 12 & 11 & 51 & 43 & 32 & 32 & 55 & 55 & 72 & 72 & 72 & 72\\\hline
H2\_JW\_sto3g & 4 & 12 & 12 & 56 & 48 & 46 & 46 & 56 & 56 & 81 & 81 & 111 & 111\\\hline
H2\_P\_631g & 6 & 152 & 146 & 417 & 287 & 324 & 324 & 199 & 199 & 223 & 223 & 223 & 223\\\hline
H4\_P\_sto3g & 6 & 158 & 152 & 409 & 293 & 316 & 316 & 210 & 210 & 229 & 229 & 229 & 229\\\hline
H2\_JW\_631g & 8 & 84 & 84 & 385 & 246 & 325 & 298 & 247 & 247 & 185 & 185 & 195 & 195\\\hline
H2\_BK\_631g & 8 & 84 & 83 & 383 & 275 & 285 & 267 & 245 & 245 & 176 & 176 & 181 & 181\\\hline
H4\_BK\_sto3g & 8 & 160 & 157 & 624 & 414 & 457 & 427 & 398 & 398 & 243 & 243 & 256 & 256\\\hline
H4\_JW\_sto3g & 8 & 160 & 160 & 589 & 387 & 473 & 435 & 396 & 396 & 262 & 262 & 273 & 273\\\hline
NH\_BK\_sto3g & 12 & 640 & 637 & 3369 & 1752 & 3115 & 2260 & 1915 & 1614 & 1304 & 1304 & 863 & 863\\\hline
LiH\_BK\_sto3g & 12 & 640 & 633 & 3198 & 1741 & 2968 & 2119 & 1755 & 1559 & 1093 & 1093 & 811 & 811\\\hline
LiH\_JW\_sto3g & 12 & 640 & 640 & 3244 & 1699 & 2950 & 2193 & 1784 & 1538 & 1127 & 1127 & 844 & 844\\\hline
NH\_JW\_sto3g & 12 & 640 & 640 & 3375 & 1759 & 2854 & 2195 & 1992 & 1665 & 1264 & 1264 & 860 & 860\\\hline
H2O\_P\_sto3g & 12 & 1071 & 1051 & 4968 & 2636 & 4532 & 3431 & 2625 & 2360 & 1816 & 1816 & 1228 & 1228\\\hline
BeH2\_P\_sto3g & 12 & 1075 & 1054 & 4926 & 2698 & 5017 & 3604 & 2779 & 2380 & 1756 & 1756 & 1260 & 1260\\\hline
CH2\_P\_sto3g & 12 & 2093 & 2054 & 8690 & 4768 & 8069 & 6027 & 4927 & 4420 & 3010 & 3010 & 2241 & 2241\\\hline
H2O\_JW\_sto3g & 14 & 1000 & 1000 & 5949 & 2842 & 5141 & 3415 & 3020 & 2370 & 2244 & 2244 & 1242 & 1242\\\hline
H2O\_BK\_sto3g & 14 & 1000 & 995 & 6141 & 2969 & 5103 & 3396 & 3253 & 2453 & 2263 & 2263 & 1254 & 1254\\\hline
CH2\_BK\_sto3g & 14 & 1488 & 1479 & 8641 & 4182 & 7963 & 5139 & 4630 & 3410 & 3057 & 3057 & 1752 & 1752\\\hline
BeH2\_JW\_sto3g & 14 & 1488 & 1488 & 8573 & 4070 & 8090 & 5334 & 4351 & 3364 & 2846 & 2846 & 1757 & 1757\\\hline
BeH2\_BK\_sto3g & 14 & 1488 & 1466 & 8647 & 4138 & 7320 & 4935 & 4332 & 3361 & 2855 & 2855 & 1695 & 1695\\\hline
CH2\_JW\_sto3g & 14 & 1488 & 1488 & 8497 & 4083 & 6820 & 4815 & 4518 & 3379 & 3105 & 3105 & 1776 & 1776\\\hline
H4\_P\_631g & 14 & 2902 & 2868 & 14739 & 7296 & 14607 & 9320 & 8321 & 6075 & 5172 & 5172 & 3124 & 3124\\\hline
H4\_JW\_631g & 16 & 1440 & 1440 & 9981 & 4373 & 7496 & 4910 & 4790 & 3784 & 2909 & 2909 & 1829 & 1829\\\hline
H4\_BK\_631g & 16 & 1440 & 1429 & 10094 & 4429 & 7743 & 4919 & 4529 & 3515 & 2890 & 2890 & 1759 & 1759\\\hline
NH3\_JW\_sto3g & 16 & 2340 & 2340 & 16032 & 6891 & 13723 & 8442 & 8824 & 6709 & 5029 & 5029 & 2692 & 2692\\\hline
NH3\_BK\_sto3g & 16 & 2340 & 2329 & 15941 & 6854 & 13317 & 8349 & 7757 & 5676 & 5033 & 5033 & 2661 & 2661\\\hline
HCl\_JW\_sto3g & 20 & 684 & 684 & 5946 & 2484 & 4742 & 3050 & 3205 & 2557 & 2669 & 2248 & 1159 & 1159\\\hline
HCl\_BK\_sto3g & 20 & 684 & 682 & 5233 & 2383 & 4590 & 3015 & 3490 & 2743 & 2586 & 2150 & 1156 & 1156\\\hline

    \end{tabular}
    \end{changemargin}
    \caption{Square connectivity table for molecular data experiment.}
    \label{table:squareconnmolecules}
\end{table}

\end{document}